\documentclass{article}
\usepackage{epsfig}
\usepackage{color}
\usepackage{amsmath}
\usepackage{amssymb}
\usepackage{lscape}
\usepackage{graphicx}

\makeatletter
\@addtoreset{equation}{section}

\begin{document}

%\fontsize{14pt}{0.8cm}\selectfont
\fontsize{11pt}{0.5cm}\selectfont

\begin{center}

\vspace*{4cm}
{
\huge

Wilson loops of Chern-Simons-matter Theories

\vspace{5mm}

with one parameter and Picard-Fuchs equations

}

\vspace{2cm}

{
\Large 
Takao Suyama\footnote{suyama.takao@nihon-u.ac.jp}
}

\vspace{1cm}

{
\Large

College of Engineering, Nihon University, 

\vspace{5mm}

Fukushima 963-8642, Japan
}

\vspace{2cm}

{\bf Abstract}
\end{center}

\vspace{5mm}

Extending our previous work, we investigate Wilson loops of flavored Chern-Simons theory and ABJ theory by the use of Picard-Fuchs equations. 
The singularity structure of the equations is analyzed in detail, and local solutions around the singularities are determined. 
Each theory has one parameter, and the parameter dependence of the Wilson loops is also determined. 
In flavored Chern-Simons theory, we find an interesting phase transition. 
For $n$-times-wound Wilson loops, we determine their $n$-dependence by using a recursion relation obtained in our previous work. 

\newpage

\tableofcontents

\vspace{1cm}

\section{Introduction}

\vspace{5mm}

In our previous work \cite{Suyama:2025ydf}, we investigated Wilson loops in ${\cal N}=3$ pure Chern-Simons (CS) theory and ABJM theory \cite{Aharony:2008ug} in the planar limit. 
Based on the localization formula \cite{Kapustin:2009kz}, we can regard these theories as matrix models. 
The planar resolvents for the matrix models can be determined \cite{Aganagic:2002wv}\cite{Halmagyi:2003ze}. 
It was found in \cite{Suyama:2025ydf}\cite{Suyama:2016nap} that the derivative of the resolvent defines a simple algebraic curve. 
The vevs of the Wilson loops and the 't~Hooft couplings can be written as integrals on the algebraic curve. 
Using this fact, we derived the Picard-Fuchs (PF) equations, a set of linear differential equations, for these quantities. 
Physically interesting properties of the theory, for example the weak and strong coupling limits, are encoded into the regular singularities of the equations. 
In this paper, we extend the analysis in \cite{Suyama:2025ydf} to one-parameter deformations of those theories which are briefly reviewed below. 
The PF equations obtained from the resolvent, not its derivative, were discussed, for example, in \cite{Aganagic:2002wv}\cite{Marino:2009jd} based on relations to topological string theory. 
Our analysis does not rely on such  relations, and therefore, we expect that it would have a wider range of applications. 
For a review on Wilson loops of Chern-Simons-matter theories, see for example \cite{Drukker:2019bev}. 

${\cal N}=3$ CS theory with gauge group ${\rm U}(N)$ can be deformed by adding $N_f$ massless hypermultiplets in the fundamental representation. 
The ${\cal N}=3$ supersymmetry determines the superpotential of this theory completely \cite{Gaiotto:2007qi}\cite{Suyama:2010hr}. 
We call this flavored Chern-Simons (fCS) theory. 
One interesting property of fCS theory is that the vacuum of this theory preserves the supersymmetry as long as 
\begin{equation}
N\ \le\ k+N_f
   \label{SUSY}
\end{equation}
is satisfied, where $k>0$ is the Chern-Simons level. 
The simplest case $N_f=0$ with ${\cal N}=1$ supersymmetry was investigated in \cite{Witten:1999ds}\footnote{
In fact, \cite{Witten:1999ds} discusses ${\cal N}=1$ pure Yang-Mills theory with Chern-Simons term whose low energy effective theory is ${\cal N}=1$ CS theory. 
}, in which $k$ in (\ref{SUSY}) is replaced with $2k$. 
This was extended to ${\cal N}=2,3$ CS theories in \cite{Bergman:1999na}\cite{Ohta:1999iv} by using the brane construction \cite{Kitao:1998mf}. 
The inclusion of fundamental matters was discussed in \cite{Giveon:2008zn} for ${\cal N}=2$ theory and in \cite{Suyama:2011yz} for ${\cal N}=3$ theory. 
These theories are simple examples in which physical properties of a theory may change qualitatively according to the value of a parameter. 

ABJ theory \cite{Hosomichi:2008jb}\cite{Aharony:2008gk} has the gauge group ${\rm U}(N_1)\times{\rm U}(N_2)$. 
This reduces to ABJM theory when $N_1=N_2$. 
Then, the difference $N_1-N_2$ can be regarded as the deformation parameter. 
In the gravity dual, this parameter is related to a background flux \cite{Aharony:2008gk}. 
The parameter space of ABJ theory was investigated in \cite{Drukker:2010nc}. 
It is argued in \cite{Aharony:2008gk} that these parameters should be restricted to the range 
\begin{equation}
|N_1-N_2|\ \le\ |k|, 
\end{equation}
where $k$ is the Chern-Simons level of the ${\rm U}(N_1)$ factor, in order for the theory to be consistent as a unitary theory. 
Note that this theory has two coupling constants $N_1/k$ and $N_2/k$ in the planar limit. 
This means that if $N_1-N_2$ is comparable to $k$, then at least one of the coupling constants cannot become small, and therefore, we cannot employ the perturbative method for calculating physical quantities of interest. 

\vspace{5mm}

In this paper, we investigate these two theories by using the PF equations for the Wilson loops and the 't~Hooft couplings. 
Interestingly, the equations for fCS and ABJ theories share the same form, although they look quite different from each other as Chern-Simons-matter theories. 
This is due to the fact that their algebraic curves are defined by equations of a common form. 
For fCS theory, we can check the validity of our analysis since we can obtain the exact expressions for the Wilson loops and the 't~Hooft coupling directly from the planar resolvent. 

As observed in \cite{Suyama:2025ydf}, physically interesting information is encoded into the singularities of the PF equations. 
Since we find concise expressions for the equations, we can understand the structure of the singularities in detail. 
We will show that most of the singularities turn out to be simple poles. 
There also exist double poles, but they can be eliminated by a ``gauge transformation.'' 
It is then a standard procedure to determine the local solution around each singularity. 
We will see that some of the singularities correspond to characteristic points in the parameter space of the theory. 
For example, we will find an interesting phase transition in fCS theory, and a simple pole in the PF equations plays a role in the phase transition. 

As usual for differential equations, the local solutions contain integration constants. 
For equations in this paper, the constants are actually functions of the deformation parameter. 
Interestingly, we have another set of differential equations which are also satisfied by the Wilson loops and the 't~Hooft couplings. 
We use these equations to determine the parameter dependence of the integration constants, up to parameter-independent integration constants. 
In many cases, the remaining constants can be finally fixed by requiring that the solution should reproduce results for CS or ABJM theories when the deformation parameter is turned off. 

It turns out that a geometric viewpoint based on the algebraic curve is quite helpful for understanding the singularity structure of the PF equations. 
We observe that the singularities are related to the various degenerations of the algebraic curve. 
The structure of the residue matrix at each singularity seems to be related to the complexity of the corresponding degeneration. 
The gauge transformation mentioned above can be deduced from a procedure which simplifies the degeneration corresponding to the double pole. 

We also studied in \cite{Suyama:2025ydf} $n$-times-wound Wilson loops for $n\ge1$ by deriving a recursion relation for them. 
This relation was rewritten as a differential equation for the generating function of the Wilson loops. 
In this paper, we use this differential equation to determine the exact $n$-dependence of the local solutions for the Wilson loops. 
The $n$-dependence of the Wilson loops was also investigated in \cite{Drukker:2010nc}. 

\vspace{5mm}

This paper is organized as follows. 
In section \ref{algebraic curve}, we define the algebraic curves for fCS and ABJ theories, and write down the integral formulas for the Wilson loops and the 't~Hooft couplings. 
Then, we derive the PF equations for them in section \ref{PF eqs}. 
The singularities of the equations are analyzed in section \ref{singularities}, and the residue matrices at the singularities are determined in section \ref{residue matrix}. 
Based on the information obtained so far, we determine the local solutions around the singularities in section \ref{local solutions}. 
In section \ref{para dependence}, we analyze the parameter dependence of the physical properties of fCS and ABJ theories, and find a phase transition in fCS theory. 
In section \ref{recursion relation}, we use the recursion relation for the multiply-wound Wilson loops to obtain their exact leading order terms. 
Section \ref{discuss} is devoted to discussion. 
In appendix \ref{app-fCS}, we derive the algebraic curve for fCS theory. 
Appendix \ref{app-WL} contains some formulas for the Wilson loops. 
In appendix \ref{app-exact}, we derive some exact results for fCS theory. 

\vspace{1cm}

\section{Algebraic curves}   \label{algebraic curve}

\vspace{5mm}

The partition function of an ${\cal N}=3$ Chern-Simons-matter theory can be reduced to an eigenvalue integral \cite{Kapustin:2009kz}. 
The vevs of some BPS operators, including Wilson loops, can be also given as similar integrals. 
In the planar limit, the information of these integrals is encoded into an algebraic curve \cite{Suyama:2025ydf}. 
In \cite{Suyama:2025ydf}, we investigated ${\cal N}=3$ CS theory and ABJM theory. 
We consider their one-parameter deformations in the following. 

\vspace{5mm}

First, we consider fCS theory which has the gauge group ${\rm U}(N)$ with the Chern-Simons level $k$, and $N_f$ massless hypermultiplets in the fundamental representation of ${\rm U}(N)$. 
In the planar limit, we take $N,k,N_f$ to become large while keeping 
\begin{equation}
t\ :=\ \frac{2\pi iN}k, \hspace{1cm} \nu\ :=\ \frac{2\pi iN_f}k
   \label{parameters-fCS}
\end{equation}
fixed, where $t$ is the 't~Hooft coupling. 

The algebraic curve for fCS theory is defined by 
\begin{equation}
y^2\ =\ \frac{(x-2)(x-\beta)^2}{(x+2)^2(x-\alpha)}. 
   \label{fCS curve}
\end{equation}
The derivation of this algebraic curve is summarized in appendix \ref{app-fCS}. 
The parameters $\alpha$ and $\beta$ are given in terms of a complex variable $\gamma$ as 
\begin{equation}
\alpha\ :=\ \gamma^2-2, \hspace{1cm} \beta\ :=\ -\frac\nu2\gamma-2. 
\end{equation}
As we will see shortly, $\gamma$ is related to $t$. 
Originally, $\gamma$ dictates the position of the branch cut of the resolvent which describes the eigenvalue distribution. 
The limit $t\to0$ corresponds to $\gamma\to2$. 
On the other hand, the limit $t\to\infty$ depends on the value of $\nu$, as we will show in section \ref{para dependence}. 

\vspace{5mm}

The second theory we consider is ABJ theory. 
The gauge group is ${\rm U}(N_1)\times{\rm U}(N_2)$ whose Chern-Simons levels are $k$ and $-k$, respectively. 
As in \cite{Suyama:2025ydf}, we take $N_1,N_2,k$ to become large while keeping 
\begin{equation}
t_1\ :=\ \frac{2\pi iN_1}k, \hspace{1cm} t_2\ :=\ \frac{2\pi iN_2}k
\end{equation}
fixed. 
ABJM theory corresponds to $t_1=t_2$, and therefore $B:=t_1-t_2$ is the deformation parameter. 
The algebraic curve for ABJ theory is defined by \cite{Suyama:2025ydf} 
\begin{equation}
y^2\ =\ \frac{(x-2)(x+2)}{(x-\alpha)(x-\beta)}, 
   \label{ABJ curve}
\end{equation}
where $\alpha$ and $\beta$ have a simple relation to $t_1$ and $t_2$ as \cite{Marino:2009jd}\cite{Suyama:2025ydf} 
\begin{equation}
t_1-t_2\ =\ \log\frac{\alpha-\beta}4. 
\end{equation}
This suggests that we should choose the following parametrization 
\begin{equation}
\alpha\ :=\ \gamma+2e^B, \hspace{1cm} \beta\ :=\ \gamma-2e^B. 
\end{equation}
The parameter $B$ is related to a background flux in the gravity dual \cite{Aharony:2008gk}. 

For a non-zero $B$, since $t_1$ and $t_2$ cannot vanish simultaneously, we can take either $t_1\to0$ or $t_2\to0$. 
These limits correspond to the limits $\gamma\to2-2e^B$ or $\gamma\to-2+2e^B$, respectively. 
Therefore, the results for these limits should contain some non-perturbative results. 
Note that these two limits degenerate to $\gamma\to0$ for ABJM theory. 
On the other hand, the strong coupling limit corresponds to $\gamma\to\infty$ for any $B$, as we will see in section \ref{para dependence}. 

\vspace{5mm}

We find that the algebraic curves for fCS and ABJ theories are defined by similar equations (\ref{fCS curve})(\ref{ABJ curve}). 
Therefore, they can be analyzed simultaneously by considering an algebraic curve defined by 
\begin{equation}
y^2\ =\ \prod_{i=1}^4(x-\alpha_i)^{l_i}, 
   \label{general curve}
\end{equation}
where $\alpha_i$ are assumed to depend on $\gamma$. 
For fCS and ABJ theories, we have 
\begin{equation}
(\alpha_1,\alpha_2,\alpha_3,\alpha_4)\ =\ (2,-2,\alpha,\beta). 
   \label{restricted}
\end{equation}
The exponents $l_i$ are given as 
\begin{equation}
(l_1,l_2,l_3,l_4)\ =\ \left\{
\begin{array}{cc}
(1,-2,-1,2), & (\mbox{fCS}) \\ [2mm]
(1,1,-1,-1). & (\mbox{ABJ})
\end{array}
\right.
\end{equation}
Note that \(\sum_{i=1}^4l_i=0\) holds in both theories. 
In the following, we assume that the sum always vanishes. 

The $x$-derivative on both sides of (\ref{general curve}) gives 
\begin{equation}
2\partial_xy\ =\ \sum_{i=1}^4\frac{l_iy}{x-\alpha_i}. 
   \label{x-derivative}
\end{equation}
This equation plays important roles in the following. 

\vspace{5mm}

We are interested in the Wilson loops in the theories, to which the localization can apply. 
The vevs of these Wilson loops can be given as contour integrals on the algebraic curve \cite{Suyama:2025ydf}. 
Explicitly, the vev of $n$-times-wound Wilson loop for fCS theory is given as 
\begin{equation}
tW_n\ =\ -\frac1{2n}\int_C\frac{dz}{2\pi i}z^{n-1}y(x), \hspace{1cm} x\ =\ z+z^{-1}, 
   \label{z-integral-WL}
\end{equation}
where the integration contour $C$ is given in appendix \ref{app-fCS}. 
The Wilson loops for ABJ theory are given similarly. 
Note that ABJ theory has two sets of Wilson loops corresponding to the ${\rm U}(N_1)$ and ${\rm U}(N_2)$ factors of the gauge group. 
We denote them as $W^{(1)}_n$ and $W^{(2)}_n$, respectively. 
They are distinguished by different choices of the contour $C$, as explained in \cite{Suyama:2025ydf}. 

In fact, $tW_n$ are not convenient for the following analysis. 
Instead, as in \cite{Suyama:2025ydf}, we introduce 
\begin{equation}
\pi_n\ :=\ \int_C\frac{dx}{2\pi i}x^ny(x), 
   \label{integral-WL}
\end{equation}
where the contour $C$ mapped to the $x$-plane is also denoted by $C$. 
For example, we have $\pi_0=-4tW_1$. 
In general, $tW_n$ are linear combinations of $\pi_m$. 
See appendix \ref{app-WL} for a general formula. 

In order to regard $W_n$ as a function of $t$, we need the following integral formula \cite{Suyama:2025ydf} 
\begin{equation}
t\ =\ -\frac1{4\pi i}\int_C\frac{dz}z\log z\ y(x), 
   \label{integral-tH}
\end{equation}
which relates $t$ to $\gamma$. 
This can be used to eliminate $\gamma$ from $W_n$. 
We also define 
\begin{equation}
\tau_n\ :=\ -\frac1{4\pi i}\int_C\frac{dz}z\log z\ x^ny(x), \hspace{1cm} \tau_0\ =\ t, 
\end{equation}
which will become necessary later. 
Note that $\tau_n$ with $n\ge1$ are just auxiliary quantities for us, and we will not discuss their physical meaning in this paper. 

\vspace{1cm}

\section{Picard-Fuchs equations}   \label{PF eqs}

\vspace{5mm}

We defined $\pi_n$ and $\tau_n$ as integrals of $y(x)$. 
This implies that the $\gamma$-dependence of them can be understood from the $\gamma$-dependence of $y(x)$ which is determined by the defining equation (\ref{general curve}) of the algebraic curve. 
In this section, we will translate the information contained in the algebraic curve into a set of differential equations for $\pi_n$ and $\tau_n$. 

\vspace{5mm}

First, we consider the $\gamma$-dependence of $\pi_n$. 
We start with 
\begin{equation}
2\partial_\gamma y\ =\ \sum_{i=1}^4\partial_\gamma\alpha_i\frac{-l_iy}{x-\alpha_i}, 
   \label{gamma-derivative}
\end{equation}
which is obtained from (\ref{general curve}). 
By integrating both sides, we obtain an equation for $\partial_\gamma\pi_0$. 
We also obtain 
\begin{eqnarray}
2\partial_\gamma (xy) 
&=& -\sum_{i=1}^4l_i\partial_\gamma\alpha_i\,y+\sum_{i=1}^4\alpha_i\partial_\gamma\alpha_i\frac{-l_iy}{x-\alpha_i}, 
   \label{gamma-derivative 2} \\
2\partial_\gamma(x^2y) 
&=& -\sum_{i=1}^4l_i\partial_\gamma\alpha_i\,xy-\sum_{i=1}^4l_i\alpha_i\partial_\gamma\alpha_i\,y+\sum_{i=1}^4(\alpha_i)^2\partial_\gamma\alpha_i\frac{-l_iy}{x-\alpha_i}, 
   \label{gamma-derivative 3}
\end{eqnarray}
and so on, by multiplying $x^n$ to (\ref{gamma-derivative}). 
These relations give equations for $\partial_\gamma\pi_1$ and $\partial_\gamma\pi_2$. 

Suppose that $-l_iy/(x-\alpha_i)$ can be written as linear combinations of $x^ny$, up to a total derivative term with respect to $x$. 
Then, we obtain a set of linear differential equations for $\pi_n$ by integrating both sides of the above equations with respect to $x$ along a closed contour $C$. 

For this purpose, we define polynomials $f_i(x)$ as 
\begin{equation}
f_i(x)\ :=\ \frac{f(x)}{x-\alpha_i}, \hspace{1cm} f(x)\ :=\ \prod_{i=1}^4(x-\alpha_i), 
   \label{f_i,f}
\end{equation}
and multiply them to the equation (\ref{x-derivative}). 
Then we obtain the desired expressions  
\begin{equation}
\frac{-l_iy}{x-\alpha_i}\ =\ \frac1{f_i(\alpha_i)}p_i(x)y-\frac2{f_i(\alpha_i)}\partial_x\Bigl( f_i(x)y \Bigr), 
   \label{constraint}
\end{equation}
where $p_i(x)$ are polynomials of degree 2 defined as 
\begin{equation}
p_i(x)\ :=\ 2f_i'(x)+\sum_{j=1}^4l_j\frac{f_i(x)-f_i(\alpha_j)}{x-\alpha_j}. 
\end{equation}
Note that $f_i(\alpha_j)=0$ hold for $i\ne j$. 

In order to write down the equations (\ref{gamma-derivative})(\ref{gamma-derivative 2})(\ref{gamma-derivative 3}) in a compact form, we introduce 
\begin{equation}
v(x)\ :=\ 
\left[
\begin{array}{c}
1 \\
x \\
x^2
\end{array}
\right]
, \hspace{1cm} M(x)\ :=\ 
\left[
\begin{array}{ccc}
0 & 0 & 0 \\
1 & 0 & 0 \\
x & 1 & 0
\end{array}
\right]. 
\end{equation}
We also define row vectors $p_i$ such that $p_iv(x)=p_i(x)$ hold. 
Then, we can combine the equations (\ref{gamma-derivative})(\ref{gamma-derivative 2})(\ref{gamma-derivative 3}) into the form 
\begin{equation}
2\partial_\gamma\Bigl( v(x)y \Bigr)\ =\ \sum_{i=1}^4\partial_\gamma\alpha_i\left[ -l_iM(\alpha_i)+\frac1{f_i(\alpha_i)}v(\alpha_i)p_i \right]v(x)y+\partial_x(\cdots), 
\end{equation}
where we have used the expressions (\ref{constraint}). 
By integration, we obtain 
\begin{equation}
\partial_\gamma\Pi\ =\ A\Pi, \hspace{1cm} \Pi\ :=\ \int_C\frac{dx}{2\pi i}v(x)y\ =\ 
\left[
\begin{array}{c}
\pi_0 \\
\pi_1 \\
\pi_2
\end{array}
\right], 
   \label{PF-WL}
\end{equation}
where $A$ is the $3\times3$ matrix defined as 
\begin{equation}
A\ :=\ \frac12\sum_{i=1}^4\partial_\gamma\alpha_i\left[ -l_iM(\alpha_i)+\frac1{f_i(\alpha_i)}v(\alpha_i)p_i \right]. 
   \label{A}
\end{equation}
We call this system of first-order linear differential equations the PF equation for Wilson loops. 

Note that $\pi_n$ with $n\ge3$ may also satisfy similar differential equations. 
As will be discussed in section \ref{recursion relation}, it is more convenient to employ a recursion relation for $\pi_n$. 
By using this, we can determine all $\pi_n$ in terms of $\pi_0,\pi_1$ and $\pi_2$ obtained from (\ref{PF-WL}). 

\vspace{5mm}

Another set of differential equations for $\tau_n$ can be also derived as follows. 
We multiply $\log z$ to (\ref{gamma-derivative})(\ref{gamma-derivative 2})(\ref{gamma-derivative 3}), where $z$ is related to $x$ as $x=z+z^{-1}$. 
Then we obtain 
\begin{equation}
2\partial_\gamma\Bigl( \log z\,v(x)y \Bigr)\ =\ -\sum_{i=1}^4l_i\partial_\gamma\alpha_iM(\alpha_i)\log z\,v(x)y+\sum_{i=1}^4\partial_\gamma\alpha_iv(\alpha_i)\frac{-l_i\log z\,y}{x-\alpha_i}. 
\end{equation}
By applying the integration $-\frac1{4\pi i}\int_C\frac{dz}z$ on both sides, we obtain a system of differential equations for $\tau_n$, provided that we rewrite $-l_i\log z\,y/(x-\alpha_i)$ into a suitable form. 

For this purpose, we use 
\begin{equation}
\frac{-l_i\log z\,y}{x-\alpha_i}\ =\ \frac1{f_i(\alpha_i)}p_i\log z\,v(x)y-\frac2{f_i(\alpha_i)}\log z\,\partial_x\Bigl( f_i(x)y \Bigr), 
\end{equation}
which are obtained from (\ref{constraint}). 
The last term in the right-hand side must be further rewritten since its integration does not vanish. 

To proceed further, we restrict our analysis to the case (\ref{restricted}) which includes fCS and ABJ theories. 
Then, we only need to rewrite the above relations for $i=3,4$ since $\partial_\gamma\alpha_1=\partial_\gamma\alpha_2=0$ holds. 
For the case $i=3$, we find 
\begin{equation}
\log z\,\partial_x\Bigl( f_3(x)y \Bigr)\ =\ -z\partial_zx\,(x-\alpha_4)y+\log z\,x(x-\alpha_4)y+z\partial_z(\cdots), 
\end{equation}
where we used 
\begin{equation}
x^2-4\ =\ (z-z^{-1})^2, \hspace{1cm} z-z^{-1}\ =\ z\partial_zx. 
\end{equation}
In the right-hand side, the last term vanishes by the integration $-\frac1{4\pi i}\int_C\frac{dz}z$. 
The second term gives a linear combination of $\tau_1$ and $\tau_2$, while the first term gives a linear combination of $\pi_0$ and $\pi_1$ since 
\begin{equation}
-\frac1{4\pi i}\int_C\frac{dz}z z\partial_zx\,x^ny\ =\ -\frac12\pi_n. 
\end{equation}
A similar relation can be obtained for $i=4$. 
Now, we obtain 
\begin{equation}
\frac{-l_i\log z\,y}{x-\alpha_i}\ =\ \frac1{f_i(\alpha_i)}\tilde p_i\log z\,v(x)y+\frac2{f_i(\alpha_i)}z\partial_zx\,u_iv(x)y+z\partial_z(\cdots) 
\end{equation}
for $i=3,4$, where 
\begin{equation}
\tilde p_3\ :=\ p_3+(0,2\alpha_4,-2), \hspace{1cm} \tilde p_4\ :=\ p_4+(0,2\alpha_3,-2), 
\end{equation}
and 
\begin{equation}
u_3\ :=\ (-\alpha_4,1,0), \hspace{1cm} u_4\ :=\ (-\alpha_3,1,0). 
\end{equation}

Finally, we obtain 
\begin{equation}
\partial_\gamma T\ =\ \tilde AT+J\Pi, \hspace{1cm} T\ :=\ -\frac1{4\pi i}\int_C\frac{dz}z\log z\,v(x)y\ =\ 
\left[
\begin{array}{c}
\tau_0 \\
\tau_1 \\
\tau_2
\end{array}
\right], 
   \label{PF-tH}
\end{equation}
where 
\begin{eqnarray}
\tilde A 
&:=& \frac12\sum_{i=3,4}\partial_\gamma\alpha_i\left[ -l_iM(\alpha_i)+\frac1{f_i(\alpha_i)}v(\alpha_i)\tilde p_i \right], 
   \label{tilde A} \\ [2mm] 
J 
&:=& -\frac12\sum_{i=3,4}\frac{\partial_\gamma\alpha_i}{f_i(\alpha_i)}v(\alpha_i)u_i. 
\end{eqnarray}
We call this system of first-order linear differential equations the PF equation for 't~Hooft coupling. 
Recall that $\Pi$ can be determined independently of $T$. 
Therefore, the term $J\Pi$ can be regarded as a source term for $T$. 

\vspace{5mm}

So far, we have regarded $\gamma$ as a variable while $\nu$ in fCS theory and $B$ in ABJ theory as parameters. 
However, both of them are on the equal footing, when we regard them as parameters for the coefficients $\alpha_i$ of the algebraic curve (\ref{general curve}). 
Then, we can also obtain, for example, 
\begin{equation}
\partial_\nu\Pi\ =\ A_\nu\Pi, \hspace{1cm} A_\nu\ :=\ \frac12\sum_{i=1}^4\partial_\nu\alpha_i\left[ -l_iM(\alpha_i)+\frac1{f_i(\alpha_i)}v(\alpha_i)p_i \right] 
   \label{nu-PF-WL}
\end{equation}
for fCS theory. 

This can be used as follows. 
The general solution of (\ref{PF-WL}) is of the form 
\begin{equation}
\Pi\ =\ c_1(\nu)\Pi^{(1)}+c_2(\nu)\Pi^{(2)}+c_3(\nu)\Pi^{(3)}, 
\end{equation}
where $c_1(\nu)$ etc. are unknown functions of $\nu$. 
By requiring that this also satisfies (\ref{nu-PF-WL}), we obtain a set of differential equations for these functions. 
We will show in section \ref{local solutions} that they can be solved exactly in many cases of interest. 
Then $c_1(\nu)$ etc. are determined up to $\nu$-independent constants which can be fixed if they also appear in the exact expression for $\Pi$ at $\nu=0$ \cite{Suyama:2025ydf}. 

\vspace{1cm}

\section{Singularities}   \label{singularities}

\vspace{5mm}

The important information of the PF equations (\ref{PF-WL})(\ref{PF-tH}) is encoded into local solutions around singularities of the coefficient matrices $A$ and $\tilde A$. 
Since the entries of the matrices are rational functions of $\gamma$, the singularities are actually poles. 
In this section, we investigate the pole structure of $A$ and $\tilde A$ for cases $\gamma\in\mathbb C$ and $\gamma=\infty$ separately. 

\vspace{5mm}

\subsection{Poles in $\mathbb C$} \label{singularity-finite}

\vspace{5mm}

The expressions (\ref{A})(\ref{tilde A}) show that, as long as $\gamma$ is finite, the poles of $A$ and $\tilde A$ come only from the zeros of $f_3(\alpha_3)$ and $f_4(\alpha_4)$. 
The actual positions of the poles depend on the parametrization of $\alpha_i$ in terms of $\gamma$. 

\subsubsection{Poles for fCS theory}

\begin{figure}
\begin{center}
\includegraphics[width=70mm]{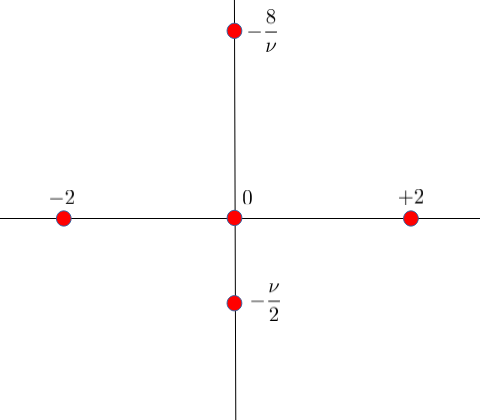}
\end{center}
\caption{
Positions of the poles of the PF equations for fCS theory. 
The value of $\nu$ is assumed to satisfy the conditions ${\rm Re}(\nu)=0$ and ${\rm Im}(\nu)\ge0$ which are physically relevant. 
}
   \label{poles-fCS}
\end{figure}

For fCS theory, the positions of the poles can be read off from 
\begin{equation}
f_3(\alpha_3)\ =\ \gamma^3(\gamma-2)(\gamma+2)\left( \gamma+\frac\nu2 \right), \hspace{1cm} f_4(\alpha_4)\ =\ -\frac{\nu^2}4\gamma^2\left( \gamma+\frac\nu2 \right)\left( \gamma+\frac8\nu \right). 
\end{equation}
These expressions imply that there may be poles at $\gamma=0,\pm2,-\frac\nu2,-\frac8\nu$. 
By expanding $A$ and $\tilde A$, we find that $\gamma=0$ is a double pole, and the others are simple poles. 
The configuration of these poles is depicted in Figure \ref{poles-fCS}. 
The local solutions around these poles will be determined in section \ref{local solutions}. 
As is well-known, it is not straightforward to determine the local solution around a double pole. 
Fortunately, the double pole at $\gamma=0$ can be eliminated by a suitable transformation of $\Pi$ and $T$, as we will show below. 

As explained in appendix \ref{app-fCS}, $\gamma$ is related to a branch point $z=a$ of the resolvent $v(z)$ as $\gamma=a^{1/2}+a^{-1/2}$. 
Then, the poles correspond to the branch points as 
\begin{equation}
\begin{array}{lcl}
\gamma\ =\ 0 & \Leftrightarrow & a\ =\ e^{\pi i}, \\ [1mm] 
\gamma\ =\ 2 & \hspace{5mm} \Leftrightarrow \hspace{5mm} & a\ =\ 1, \\ [1mm] 
\gamma\ =\ -2 & \Leftrightarrow & a\ =\ e^{2\pi i}, 
\end{array}
\end{equation}
and so on. 
For $a=1$, the branch cut of $v(z)$ shrinks to a point. 
This corresponds to the weak coupling limit $t\to0$. 
It is interesting to notice that $a=1$ and $a=e^{2\pi i}$ correspond to different points in the $\gamma$-plane. 
In fact, we will see in section \ref{para dependence} that the latter is the boundary point of the region in which the supersymmetry is preserved. 

The definition (\ref{f_i,f}) of $f_i(x)$ implies that some of $\alpha_i$ coincide among them at the poles. 
For a generic value of $\nu$, we find 
\begin{equation}
\begin{array}{lcl}
\gamma\ =\ 0 & \hspace{5mm} \Leftrightarrow \hspace{5mm} & \alpha_2\ =\ \alpha_3\ =\ \alpha_4, \\ [1mm]
\gamma\ =\ \pm2 & \Leftrightarrow & \alpha_1\ =\ \alpha_3, \\ [1mm] 
\gamma\ =\ -\frac\nu2 & \Leftrightarrow & \alpha_3\ =\ \alpha_4, \\ [1mm] 
\gamma\ =\ -\frac8\nu & \Leftrightarrow & \alpha_1\ =\ \alpha_4. 
\end{array}
\end{equation}
It is curious to notice that, at the double pole $\gamma=0$, three of $\alpha_i$ coincide among them, while only two of $\alpha_i$ coincide at the other simple poles. 
Geometrically, these correspond to various degenerations of the algebraic curve (\ref{general curve}). 

\subsubsection{Poles for ABJ theory}

\begin{figure}
\begin{center}
\includegraphics[width=70mm]{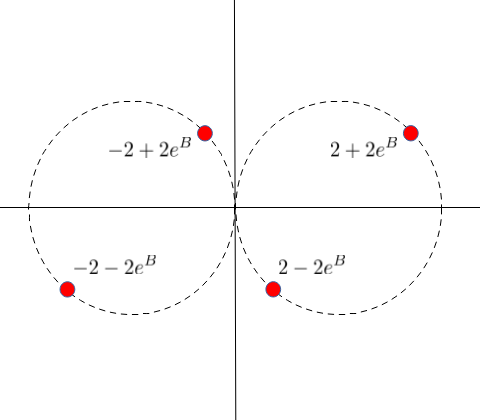}
\end{center}
\caption{
Poles of the PF equations for ABJ theory. 
The value of $B$ is assumed to satisfy the condition ${\rm Re}(B)=0$ which is physically relevant. 
}
   \label{poles-ABJ}
\end{figure}

For ABJ theory, we find 
\begin{eqnarray}
f_3(\alpha_3) 
&=& 4e^B(\gamma-2+2e^B)(\gamma+2+2e^B), \\ [1mm] 
f_4(\alpha_4) 
&=& -4e^B(\gamma-2-2e^B)(\gamma+2-2e^B). 
\end{eqnarray}
These expressions show that there may be poles at $\gamma=\pm2\pm2e^B$. 
All of them turn out to be simple poles. 
The configuration of these poles is depicted in Figure \ref{poles-ABJ}. 

Recall that ABJM theory corresponds to $B=0$. 
For this case, two poles at $\gamma=2-2e^B$ and $\gamma=-2+2e^B$ degenerate at $\gamma=0$, at which two 't~Hooft couplings $t_1,t_2$ vanish simultaneously. 
For general $B$, on the other hand, only one of the 't~Hooft couplings can vanish at each pole, and the other one remains finite. 
Therefore, local solutions around these poles contain results for finite 't Hooft coupling. 

We will see in section \ref{para dependence} that ABJ theory with $B=\pm2\pi i$ is different from ABJM theory, although these theories share the same configuration of poles. 
As reviewed in the introduction, it is argued in \cite{Aharony:2008gk} that ABJ theory is consistent as a unitary theory only when $|N_1-N_2|\ \le\ |k|$ is satisfied. 
In section \ref{para dependence}, we will find that the endpoints of this region correspond to $B=\pm2\pi i$. 

As for fCS theory, the simple poles correspond to degenerations of the algebraic curve. 
For a generic value of $B$, we find 
\begin{equation}
\begin{array}{lcl}
\gamma\ =\ 2+2e^B & \hspace{5mm} \Leftrightarrow \hspace{5mm} & \alpha_1\ =\ \alpha_4, \\ [1mm] 
\gamma\ =\ 2-2e^B & \Leftrightarrow & \alpha_1\ =\ \alpha_3, \\ [1mm] 
\gamma\ =\ -2+2e^B & \Leftrightarrow & \alpha_2\ =\ \alpha_4, \\[1mm] 
\gamma\ =\ -2-2e^B & \Leftrightarrow & \alpha_2\ =\ \alpha_3. 
\end{array}
\end{equation}
We notice that only two of $\alpha_i$ coincide at the simple poles. 
This is also the case for ABJM theory since $\alpha_1\ne\alpha_2$. 

\vspace{5mm}

\subsection{Pole at infinity}   \label{pole at infinity}

\vspace{5mm}

The PF equations (\ref{PF-WL})(\ref{PF-tH}) may also have a singularity at $\gamma=\infty$. 
In order to analyze this singularity, we introduce a new variable $\xi:=\gamma^{-1}$, in terms of which we obtain 
\begin{equation}
\partial_\xi\Pi\ =\ -\frac1{\xi^2}A(\xi^{-1})\Pi, \hspace{1cm} \partial_\xi T\ =\ -\frac1{\xi^2}\tilde A(\xi^{-1})T-\frac1{\xi^2}J(\xi^{-1})\Pi. 
\end{equation}
The singularity at $\xi=0$ is a pole. 
It is, however, more difficult to understand its origin, compared with the previous case $\gamma\in\mathbb C$. 
For example, $M(\alpha_i)$ in (\ref{A}) may also give a pole at $\xi=0$. 
A brute force calculation shows that $\xi=0$ is actually a double pole for fCS and ABJ theories. 

In order to deal with the double pole, it is helpful to recall the observation in the previous subsections that the pole structure is correlated with the degeneration of the algebraic curve. 
We found that a more complicated degeneration leads to a higher order pole. 
This suggests that a suitable modification of the algebraic curve might make the singularity of the corresponding PF equations milder than the original one. 

\vspace{5mm}

Let us examine this possibility with ABJ theory. 
We rescale the variable $x$ as $x=\gamma\hat x$. 
Then, the defining equation (\ref{ABJ curve}) becomes 
\begin{equation}
y^2\ =\ \frac{(\hat x-2\xi)(\hat x+2\xi)}{(\hat x-1-2e^B\xi)(\hat x-1+2e^B\xi)}. 
   \label{rescaled ABJ curve}
\end{equation}
This is of the general form (\ref{general curve}) with 
\begin{equation}
\alpha_1\ =\ 2\xi, \hspace{1cm} \alpha_2\ =\ -2\xi, \hspace{1cm} \alpha_3\ =\ 1+2e^B\xi, \hspace{1cm} \alpha_4\ =\ 1-2e^B\xi, 
   \label{rescaled pole-ABJ}
\end{equation}
and the same exponents $l_i$ as those of ABJ theory. 
We find that all $\alpha_i$ stay finite at $\xi=0$, and at most two of $\alpha_i$ coincide. 
We can obtain the PF equation for Wilson loops 
\begin{equation}
\partial_\xi\hat\Pi\ =\ {\cal A}\hat \Pi, \hspace{1cm} \hat\Pi\ :=\ \int_C\frac{d\hat x}{2\pi i}v(\hat x)y(\hat x)\ =\ 
\left[
\begin{array}{c}
\hat\pi_0 \\
\hat \pi_1 \\
\hat\pi_2
\end{array}
\right]. 
   \label{rescaled PF-WL}
\end{equation}
directly from the rescaled curve (\ref{rescaled ABJ curve}). 
For this case, we find 
\begin{eqnarray}
f_1(\alpha_1) 
&=& 4\xi\left( 1-(2-2e^B)\xi \right)\left( 1-(2+2e^B)\xi \right), \\ [1mm] 
f_2(\alpha_2) 
&=& -4\xi\left( 1+(2-2e^B)\xi \right)\left( 1+(2+2e^B)\xi \right), \\ [1mm]
f_3(\alpha_3) 
&=& 4e^B\xi\left( 1-(2-2e^B)\xi \right)\left( 1+(2+2e^B)\xi \right), \\ [1mm] 
f_4(\alpha_4) 
&=& -4e^B\xi\left( 1+(2-2e^B)\xi \right)\left( 1-(2+2e^B)\xi \right). 
\end{eqnarray}
These expressions imply that indeed $\cal A$ has at most a simple pole at $\xi=0$, as expected. 

Of course, this calculation is useless unless there is a relation between (\ref{rescaled PF-WL}) and the original equation (\ref{PF-WL}). 
Interestingly, it turns out that there is a simple relation. 
To see this, notice that 
\begin{equation}
\pi_n\ =\ \gamma^{n+1}\int_C\frac{d\hat x}{2\pi i}\hat x^ny\ =\ \gamma^{n+1}\hat \pi_n 
\end{equation}
hold. 
These indicate that $\hat \Pi$ defined above is related to $\Pi$ as 
\begin{equation}
\Pi\ =\ U\hat \Pi, \hspace{1cm} U\ :=\ {\rm diag}(\gamma,\gamma^2,\gamma^3). 
   \label{gauge-ABJ}
\end{equation}
The rescaled equation (\ref{rescaled PF-WL}) then implies 
\begin{equation}
\partial_\gamma\Pi\ =\ -\xi^2\left[ U{\cal A}U^{-1}+\partial_\xi U\,U^{-1} \right]\Pi. 
\end{equation}
This turns out to coincide with the original equation (\ref{PF-WL}). 
That is, the rescaling of the algebraic curve we have done above amounts to a gauge transformation by $U$ which brings $A$ to $\cal A$. 

It is not possible, however, to obtain the PF equation for 't~Hooft coupling directly from the rescaled curve (\ref{rescaled ABJ curve}) through the procedure explained in section \ref{PF eqs}. 
This is because we used a special choice $\alpha_1=2,\ \alpha_2=-2$ in the derivation, which is not available in the rescaled case (\ref{rescaled pole-ABJ}). 
Instead, we can try to perform the gauge transformation of $T$ by the same $U$. 
Then we obtain 
\begin{equation}
\partial_\xi \hat T\ =\ \tilde {\cal A}\hat T+{\cal J}\hat \Pi, \hspace{1cm} \hat T\ :=\ U^{-1}T, 
   \label{rescaled PF-tH}
\end{equation}
where 
\begin{eqnarray}
\tilde {\cal A} 
&=& -\frac1{\xi^2}U^{-1}\tilde A(\xi^{-1})U+\partial_\xi U^{-1}\,U, \\
{\cal J} 
&=& -\frac1{\xi^2}U^{-1}J(\xi^{-1})U. 
\end{eqnarray}
Remarkably, we find that $\tilde {\cal A}$ also has at most a simple pole at $\xi=0$. 

\vspace{5mm}

The same rescaling also works for fCS theory, by which we can eliminate the double pole of $A$ and $\tilde A$ at $\xi=0$. 
In fact, there is a better procedure which includes a shift of $x$. 
We choose a new coordinate $\hat x$ such that $x=\gamma\hat x-2$ holds. 
Then, the defining equation (\ref{fCS curve}) becomes 
\begin{equation}
y^2\ =\ \frac{(\hat x-\frac4\gamma)(\hat x+\frac\nu2)^2}{\hat x^2(\hat x-\gamma)}. 
\end{equation}
This is a curve of the form (\ref{general curve}) with 
\begin{equation}
\alpha_1\ =\ \frac4\gamma, \hspace{1cm} \alpha_2\ =\ 0, \hspace{1cm} \alpha_3\ =\ \gamma, \hspace{1cm} \alpha_4\ =\ -\frac\nu2. 
   \label{rescaled pole-fCS}
\end{equation}

This change of variable turns out to correspond to the gauge transformation by 
\begin{equation}
U\ =\ 
\left[
\begin{array}{ccc}
\gamma & 0 & 0 \\
-2\gamma & \gamma^2 & 0 \\
4\gamma & -4\gamma^2 & \gamma^3
\end{array}
\right]. 
   \label{gauge-fCS}
\end{equation}
We find that $\alpha_1$ and $\alpha_2$ coincide with each other in the limit $\gamma\to\infty$. 
Since $\alpha_3$ goes to infinity in this limit, the pole structure of $\cal A$ and $\tilde {\cal A}$ is not determined solely by the zeros of $f_i(\alpha_i)$. 
Nevertheless, we find that both $\cal A$ and $\tilde {\cal A}$ have at most a simple pole at $\gamma=\infty$. 

Interestingly, if we instead consider the limit $\gamma\to0$, then the roles of $\alpha_1$ and $\alpha_3$ are exchanged, and we find that the same PF equations also have at most a simple pole at $\gamma=0$. 

\vspace{5mm}

\section{Residue matrices}   \label{residue matrix}

\vspace{5mm}

The information of $A$ and $\tilde A$ at a singularity is accumulated in the residue matrices. 
We will determine them in this section. 
This is a necessary step toward obtaining local solutions which will be done in section \ref{local solutions}. 

\vspace{5mm}

\subsection{Simple poles of $A$ and $\tilde A$}

\vspace{5mm}

First, we consider a simple pole of $A$ at $\gamma=p\in\mathbb C$. 
The matrix $A$ can be expanded as 
\begin{equation}
A\ =\ \frac1\xi\sum_{n=0}^\infty A_n\xi^n, \hspace{1cm} \xi\ :=\ \gamma-p. 
   \label{expand-A}
\end{equation}
We call $A_0$ the residue matrix of $A$ at $\gamma=p$. 
Since the expression (\ref{A}) for $A$ is simple, we can deduce the structure of $A_0$ without brute force calculations. 

As mentioned in subsection \ref{singularity-finite}, the simple pole is contained in the part 
\begin{equation}
A_s\ :=\ \frac{\partial_\gamma\alpha_3}{2f_3(\alpha_3)}v(\alpha_3)p_3+\frac{\partial_\gamma\alpha_4}{2f_4(\alpha_4)}v(\alpha_4)p_4, 
\end{equation}
where we used $\alpha_1=2$ and $\alpha_2=-2$. 
This matrix can be regarded as a linear transformation from $\mathbb C^3$ to $\mathbb C^3$. 
This form of $A_s$ indicates that its image is spanned by $v(\alpha_3)$ and $v(\alpha_4)$. 
There exists another vector $v_\perp$ which satisfy 
\begin{equation}
p_3v_\perp\ =\ p_4v_\perp\ =\ 0. 
\end{equation}
Then $v_\perp$ is annihilated by $A_s$, or in other words, $v_\perp$ evaluated at $\gamma=p$ is an eigenvector of $A_0$ belonging to the eigenvalue $0$.  

Let $\alpha_{i*}$ denote the value of $\alpha_i$ at $\gamma=p$. 
There are two cases $\alpha_{3*}\ne\alpha_{4*}$ and $\alpha_{3*}=\alpha_{4*}$. 
We will analyzed them separately below. 

\vspace{5mm}

(i) $\alpha_{3*}\ne\alpha_{4*}$

\vspace{5mm}

As long as $\alpha_{3*}$ and $\alpha_{4*}$ are different, the vectors $v(\alpha_{3*})$ and $v(\alpha_{4*})$ are linearly independent. 
We find that $p_3$ and $p_4$ are also linearly independent, implying that $v_\perp$ is uniquely determined up to normalization. 
Therefore, we can understand the property of $A_s$ around $\gamma=p$ from its action on the vectors $v(\alpha_3), v(\alpha_4), v_\perp$ which span $\mathbb C^3$. 
We find 
\begin{eqnarray}
A_sv(\alpha_3) 
&=& \frac12\partial_\gamma\alpha_3\left[ \frac{l_1+l_3+2}{\alpha_3-2}+\frac{l_2+l_3+2}{\alpha_3+2}+\frac{l_3+l_4+2}{\alpha_3-\alpha_4} \right] v(\alpha_3) \nonumber \\ [1mm] 
& & +\frac12\partial_\gamma\alpha_4\left[ -\frac{(l_3+2)(\alpha_3+2)}{4(\alpha_4-2)}+\frac{(l_3+2)(\alpha_3-2)}{4(\alpha_4+2)}+\frac{l_3+l_4+2}{\alpha_4-\alpha_3} \right]v(\alpha_4), \\ [1mm] 
A_sv(\alpha_4) 
&=& \frac12\partial_\gamma\alpha_3\left[ -\frac{(l_4+2)(\alpha_4+2)}{4(\alpha_3-2)}+\frac{(l_4+2)(\alpha_4-2)}{4(\alpha_3+2)}+\frac{l_3+l_4+2}{\alpha_3-\alpha_4} \right]v(\alpha_3) \nonumber \\ [1mm] 
& & +\frac12\partial_\gamma\alpha_4\left[ \frac{l_1+l_4+2}{\alpha_4-2}+\frac{l_2+l_4+2}{\alpha_4+2}+\frac{l_3+l_4+2}{\alpha_4-\alpha_3} \right]v(\alpha_4), \\ [1mm] 
A_sv_\perp 
&=& 0. 
\end{eqnarray}
Here, we only used $\alpha_1=2$ and $\alpha_2=-2$. 
We can read off the residue matrix $A_0$ of $A$ from these equations. 
Note that we can rewrite, for example, 
\begin{equation}
\frac12\partial_\gamma\alpha_3\frac{l_1+l_3+2}{\alpha_3-2}\ =\ \frac{l_1+l_3+2}2\partial_\gamma\log(\alpha_3-2). 
\end{equation}
This indicates that the entries of $A_0$ are determined by the exponents $l_i$ and the way two of $\alpha_i$ approach each other. 

For example, consider the case $p=2-2e^B$ for ABJ theory. 
We find 
\begin{equation}
A_0v(\alpha_{3*})\ =\ v(\alpha_{3*}), \hspace{1cm} A_0v(\alpha_{4*})\ =\ \frac{e^B-1}2v(\alpha_{3*}), \hspace{1cm} A_0v_{\perp*}\ =\ 0, 
\end{equation}
where $v_{\perp*}=v_\perp|_{\gamma=p}$. 
This shows that the residue matrix $A_0$ at this pole is diagonalizable, and the eigenvalues are $1,0,0$. 
The eigenvector belonging to $1$ is $v(2)$, while the eigenvectors belonging to $0$ are $v_{\perp*}$ and a linear combination of $v(2)$ and $v(2-4e^B)$. 

We find that the other simple poles in ABJ theory also have diagonalizable residue matrices with the same eigenvalues. 
The eigenvector belonging to $1$ is either $v(2)$ or $v(-2)$, depending on the poles. 

The same is also true for the poles at $\gamma=\pm2$ in fCS theory. 
On the other hand, for the pole at $\gamma=-\frac8\nu$, we still have a diagonalizable residue matrix, but the eigenvalues are $\frac52,0,0$ due to the different choice of the exponents $l_i$ for fCS theory. 

\vspace{5mm}

(ii) $\alpha_{3*}=\alpha_{4*}$

\vspace{5mm}

The case $\alpha_{3*}=\alpha_{4*}$ occurs only at the pole $\gamma=-\frac\nu2$ in fCS theory. 
Since $v(\alpha_{3*})=v(\alpha_{4*})$ holds, the image of $A_0$ is a one-dimensional vector space spanned by $v(\alpha_{3*})$. 
For this case, we have $p_{3*}=p_{4*}$, so there are two vectors $v_{\perp*}$ and $v'_{\perp*}$, independent from each other, which are then the eigenvectors of $A_0$ belonging to $0$. 

The action of $A_0$ on $v(\alpha_{3*})$ can be read off from 
\begin{eqnarray}
A_sv(\alpha_{3}) 
&\sim& \left[ \frac12\partial_\gamma\alpha_3\frac{l_3+l_4+2}{\alpha_3-\alpha_4}+\frac12\partial_\gamma\alpha_4\frac{l_3+l_4+2}{\alpha_4-\alpha_3} \right]v(\alpha_{3*}) \nonumber \\ [1mm] 
&=& \frac{l_3+l_4+2}2\partial_\gamma\log(\alpha_3-\alpha_4)\ v(\alpha_{3*}). 
\end{eqnarray}
This implies that $v(\alpha_{3*})$ is the eigenvector of $A_0$ belonging to $\frac32$. 
Therefore, $A_0$ at the pole $\gamma=-\frac\nu2$ is again diagonalizable with the eigenvalues $\frac32,0,0$. 

\vspace{5mm}

Next, consider $\tilde A$ at $\gamma=p\in\mathbb C$. 
This has the expansion 
\begin{equation}
\tilde A\ =\ \frac1\xi\sum_{n=0}^\infty \tilde A_n\xi^n. 
\end{equation}
The residue matrix $\tilde A_0$ of $\tilde A$ can be determined similarly by investigating 
\begin{equation}
\tilde A_s\ :=\ \frac{\partial_\gamma\alpha_3}{2f_3(\alpha_3)}v(\alpha_3)\tilde p_3+\frac{\partial_\gamma\alpha_4}{2f_4(\alpha_4)}v(\alpha_4)\tilde p_4. 
\end{equation}
The action of $\tilde A_s$ on $v(\alpha_3)$ and $v(\alpha_4)$ is given as 
\begin{eqnarray}
\tilde A_sv(\alpha_3) 
&=& \frac12\partial_\gamma\alpha_3\left[ \frac{l_1+l_3+1}{\alpha_3-2}+\frac{l_2+l_3+1}{\alpha_3+2}+\frac{l_3+l_4+2}{\alpha_3-\alpha_4} \right] v(\alpha_3) \nonumber \\ [1mm] 
& & +\frac12\partial_\gamma\alpha_4\left[ -\frac{(l_3+2)(\alpha_3+2)}{4(\alpha_4-2)}+\frac{(l_3+2)(\alpha_3-2)}{4(\alpha_4+2)}+\frac{l_3+l_4+2}{\alpha_4-\alpha_3} \right]v(\alpha_4), 
\end{eqnarray}
\begin{eqnarray}
\tilde A_sv(\alpha_4) 
&=& \frac12\partial_\gamma\alpha_3\left[ -\frac{(l_4+2)(\alpha_4+2)}{4(\alpha_3-2)}+\frac{(l_4+2)(\alpha_4-2)}{4(\alpha_3+2)}+\frac{l_3+l_4+2}{\alpha_3-\alpha_4} \right]v(\alpha_3) \nonumber \\ [1mm] 
& & +\frac12\partial_\gamma\alpha_4\left[ \frac{l_1+l_4+1}{\alpha_4-2}+\frac{l_2+l_4+1}{\alpha_4+2}+\frac{l_3+l_4+2}{\alpha_4-\alpha_3} \right]v(\alpha_4). 
\end{eqnarray}
The difference from the action of $A_s$ comes from the difference between $p_i$ and $\tilde p_i$. 

For ABJ theory, we find that all the residue matrices $\tilde A_0$ of $\tilde A$ at the four simple poles are diagonalizable, and they share the same set of eigenvalues $\frac12,0,0$. 
On the other hand, for fCS theory, we find 
\begin{equation}
\tilde A_0\ \sim\ \left\{
\begin{array}{ll}
{\rm diag}(\frac12,0,0), & \hspace{5mm} (\gamma=\pm2) \\ [1mm] 
{\rm diag}(2,0,0), & \hspace{5mm} (\gamma=-\frac8\nu) \\ [1mm] 
{\rm diag}(\frac32,0,0). & \hspace{5mm} (\gamma=-\frac\nu2)
\end{array}
\right. 
\end{equation}

\vspace{5mm}

\subsection{Double poles}

\vspace{5mm}

We have shown in subsection \ref{pole at infinity} that the double pole at infinity can be eliminated by a suitable gauge transformation. 
The resulting PF equations are 
\begin{equation}
\partial_\xi\hat\Pi\ =\ {\cal A}\hat\Pi, \hspace{1cm} \partial_\xi \hat T\ =\ \tilde {\cal A}\hat T+{\cal J}\hat\Pi, \hspace{1cm} \xi\ :=\ \gamma^{-1}. 
\end{equation}
Since they have only a simple pole at $\xi=0$, the coefficient matrices $\cal A$ and $\tilde {\cal A}$ have the expansions of the form (\ref{expand-A}). 

\vspace{5mm}

First, we determine the residue matrix ${\cal A}_0$ of $\cal A$ at $\xi=0$ for ABJ theory. 
Recall that $\alpha_i$ for the rescaled curve (\ref{rescaled ABJ curve}) are given in (\ref{rescaled pole-ABJ}). 
This shows that $\alpha_i$ are kept finite in the limit $\xi\to0$. 
Therefore, ${\cal A}_0$ can be understood from the singular part 
\begin{equation}
{\cal A}_s\ :=\ \sum_{i=1}^4\frac{\partial_\xi\alpha_i}{2f_i(\alpha_i)}v(\alpha_i)p_i 
\end{equation}
of $\cal A$. 
Since we have $\alpha_{1*}=\alpha_{2*}=0$ and $\alpha_{3*}=\alpha_{4*}=1$, the image of ${\cal A}_0$ regarded as a linear transformation is spanned by $v(0)$ and $v(1)$. 
The action of ${\cal A}_s$ on them turns out to be 
\begin{equation}
{\cal A}_sv(0)\ =\ \frac2\xi v(0)+{\cal O}(\xi^0), \hspace{1cm} {\cal A}_sv(1)\ =\ {\cal O}(\xi^0). 
\end{equation}
From these equations, we find 
\begin{equation}
{\cal A}_0v(0)\ =\ 2v(0), \hspace{1cm} {\cal A}_0v(1)\ =\ 0. 
\end{equation}
We need another vector $v$ in order to form a basis with $v(0)$ and $v(1)$. 
We choose 
\begin{equation}
v\ :=\ 
\left[
\begin{array}{c}
-\frac12 \\
0 \\
\frac13
\end{array}
\right]
, \hspace{1cm} {\cal A}_sv\ =\ \frac1\xi v(1)+{\cal O}(\xi^0). 
\end{equation}
This implies 
\begin{equation}
{\cal A}_0v\ =\ v(1). 
\end{equation}
This completes the determination of ${\cal A}_0$. 
In our basis, ${\cal A}_0$ is written as 
\begin{equation}
{\cal A}_0\ =\ 
\left[
\begin{array}{ccc}
2 & 0 & 0 \\
0 & 0 & 1 \\
0 & 0 & 0
\end{array}
\right]. 
\end{equation}
Note that this has a non-trivial Jordan block. 

The residue matrix $\tilde {\cal A}_0$ of $\tilde {\cal A}$ can be also analyzed, although the form of $\tilde {\cal A}$ is not as simple as $\cal A$. 
It turns out that $\tilde {\cal A}_0$ acts as 
\begin{equation}
\tilde {\cal A}_0v(0)\ =\ 2v(0), \hspace{1cm} \tilde {\cal A}_0v(1)\ =\ v(1). 
\end{equation}
For the third basis vector $\tilde v$ we choose 
\begin{equation}
\tilde v\ :=\ 
\left[
\begin{array}{c}
-1 \\
0 \\
\frac12
\end{array}
\right], \hspace{1cm} \tilde {\cal A}_0\tilde v\ =\ v(1)+\tilde v. 
\end{equation}
Now, we find that $\tilde {\cal A}_0$ is written as 
\begin{equation}
\tilde {\cal A}_0\ =\ 
\left[
\begin{array}{ccc}
2 & 0 & 0 \\
0 & 1 & 1 \\
0 & 0 & 1
\end{array}
\right], 
\end{equation}
which has a similar structure with ${\cal A}_0$. 

\vspace{5mm}

Next, we consider fCS theory which has double poles at $\gamma=0,\infty$. 
By the gauge transformation (\ref{gauge-fCS}), we found that $\cal A$ and $\tilde {\cal A}$ have at most simple poles. 
However, since either $\alpha_1$ or $\alpha_3$ in (\ref{rescaled pole-fCS}) diverges at these poles, the analysis of the residue matrices is more complicated. 
Calculations show that, in fact, the residue matrices at the poles are diagonalizable. 
Explicitly, we find 
\begin{equation}
{\cal A}_0\ \sim\ \left\{
\begin{array}{lc}
{\rm diag}(-\frac12,-1,-3), & (\gamma=0) \\ [1mm] 
{\rm diag}(1,\frac12,-3),  & (\gamma=\infty)
\end{array}
\right.
\end{equation}
and 
\begin{equation}
\tilde {\cal A}_0\ \sim\ \left\{
\begin{array}{lc}
{\rm diag}(-1,-2,-3), & (\gamma=0) \\ [1mm] 
{\rm diag}(\frac32,1,-1).  & (\gamma=\infty)
\end{array}
\right.
\end{equation}

\vspace{1cm}

\section{Local solutions}   \label{local solutions}

\vspace{5mm}

We have observed that the PF equations (\ref{PF-WL})(\ref{PF-tH}) have only simple poles, after a gauge transformation if necessary. 
These simple poles are nothing but the regular singularities of the equations. 
The analysis of local solutions around such singularities is the standard procedure. 
See for example \cite{Haraoka}. 
In this section, we determine the local solutions around the regular singularities. 

Our main interest is the behavior of the Wilson loop $W_1$ with respect to the 't Hooft coupling $t$. 
For each regular singularity, we choose a local coordinate $\xi$ for which $\xi=0$ is the position of the singularity. 
It will turn out that the leading term of $t$ around $\xi=0$ is typically of the form 
\begin{equation}
t\ \sim\ c_t\,\xi^{\delta_1}(\log\xi)^{\delta_2}, 
   \label{exponents 1}
\end{equation}
where at least one of $\delta_1$ and $\delta_2$ is non-zero. 
By inverting this relation, we obtain 
\begin{equation}
\xi\ \sim\ \left\{
\begin{array}{lc}
\Bigl( t/c_t \Bigr)^{1/\delta_1}, & \hspace*{5mm} (\delta_1\ne0) \\ [3mm] 
\exp\Bigl[ (t/c_t)^{1/\delta_2} \Bigr]. & \hspace*{5mm} (\delta_1=0,\ \delta_2\ne0)
\end{array}
\right.
\end{equation}

As shown in appendix \ref{app-WL}, $\pi_0$ is related to $W_1$ as $\pi_0=-4tW_1$. 
We will see below that the leading term of $\pi_0$ is also of the form 
\begin{equation}
\pi_0\ \sim\ c_W\,\xi^{\delta_3}(\log\xi)^{\delta_4}
   \label{exponents 2}
\end{equation}
in many cases. 
By eliminating $\xi$, we obtain 
\begin{equation}
W_1\ \sim\ \left\{
\begin{array}{lc}
c_W'\Bigl( t/c_t \Bigr)^{\delta_3/\delta_1-1}\Bigl( \log(t/c_t) \Bigr)^{\delta_4}, & \hspace{5mm} (\delta_1\ne0) \\ [3mm] 
c_W''\Bigl( t/c_t \Bigr)^{\delta_4/\delta_2-1}\exp\Bigl[ \delta_3(t/c_t)^{1/\delta_2} \Bigr]. & \hspace{5mm} (\delta_1=0,\ \delta_2\ne0)
\end{array}
\right.
\end{equation}

An exponential growth of $W_1$ is obtained for a Chern-Simons-matter theory with a semi-classical gravity dual. 
For example, ABJM theory gives $\delta_1=0, \delta_2=2$ \cite{Marino:2009jd}, and GT theory \cite{Gaiotto:2009mv}\cite{Gaiotto:2009yz} gives $\delta_1=0, \delta_2=3$ \cite{Suyama:2011yz}. 
In this sense, the values of the exponents $\delta_a$ $(a=1,\cdots,4)$, especially $\delta_1$ and $\delta_2$, are quite important in the context of AdS/CFT correspondence \cite{Maldacena:1997re}. 
We will see that $\delta_a$ are mainly determined by the residue matrix of the regular singularity, with a shift if  a gauge transformation is necessary to eliminate a double pole. 

\vspace{5mm}

\subsection{fCS theory at $\gamma=2$}

\vspace{5mm}

First, we consider fCS theory. 
As mentioned in section \ref{algebraic curve}, the pole at $\gamma=2$ corresponds to the weak coupling limit. 
Therefore, we require that both $\Pi$ and $T$ vanish at $\gamma=2$. 
We choose a local coordinate $\xi=\gamma-2$ in this subsection. 

\vspace{5mm}

The PF equation for Wilson loops (\ref{PF-WL}) can be written as 
\begin{equation}
\xi\partial_\xi\Pi\ =\ \left( \sum_{n=0}^\infty A_n\xi^n \right)\Pi. 
   \label{PF-WL-fCS_2}
\end{equation}
Suppose that this has a solution of the form 
\begin{equation}
\Pi\ =\ \sum_{m=0}^\infty \Pi_n\xi^{n+\rho}. 
\end{equation}
Substituting this into (\ref{PF-WL-fCS_2}), we obtain 
\begin{equation}
(n+\rho)\Pi_n\ =\ \sum_{m=0}^nA_{n-m}\Pi_m
   \label{mode condition}
\end{equation}
for $n\ge0$. 
The condition for $n=0$ is simply 
\begin{equation}
\rho\Pi_0\ =\ A_0\Pi_0. 
\end{equation}
This shows that $\rho$ is an eigenvalue of the residue matrix $A_0$, and $\Pi_0$ is an eigenvector belonging to $\rho$. 
Recall that $A_0$ for this case has eigenvalues $1,0,0$. 
Let $v_1,v_2,v_3$ be the corresponding eigenvectors, respectively. 
Let us choose 
\begin{equation}
\rho\ =\ 1, \hspace{1cm} \Pi_0\ =\ v_1\ =\ 
\left[
\begin{array}{c}
1 \\
2 \\
4
\end{array}
\right]. 
\end{equation}
For $n\ge1$, the conditions (\ref{mode condition}) give the relations 
\begin{equation}
\Bigl[ (n+1)I-A_0 \Bigr]\Pi_n\ =\ \sum_{m=0}^{n-1}A_{n-m}\Pi_m, 
\end{equation}
where $I$ is the identity matrix. 
Since the matrix $(n+1)I-A_0$ is invertible for all $n\ge1$, we can determine $\Pi_n$ recursively. 
We denote the resulting solution by $\Pi^{(1)}$. 

One might try to find another solution by assuming 
\begin{equation}
\Pi^{(2)}\ \stackrel{?}{=}\ \sum_{m=0}^\infty \Pi^{(2)}_n\xi^n, \hspace{1cm} \Pi^{(2)}\ =\ v_2\ =\ 
\left[
\begin{array}{c}
1 \\
0 \\
-\frac{2\nu-4}3
\end{array}
\right]. 
\end{equation}
However, this ansatz encounters a problem at $n=1$ at which we are required to solve 
\begin{equation}
\Bigl[ I-A_0 \Bigr]\Pi^{(2)}_1\ \stackrel{?}{=}\ A_1\Pi^{(2)}_0. 
\end{equation}
Since the matrix $I-A_0$ annihilates $v_1$, the left-hand side is a linear combination of $v_2$ and $v_3$. 
Therefore, if there is a $v_1$-component in the right-hand side, then this equation does not have a solution. 
This phenomenon is known as the resonance. 

The correct ansatz turns out to be 
\begin{equation}
\Pi^{(2)}\ =\ \sum_{m=0}^\infty \Pi^{(2)}_n\xi^n+c\Pi^{(1)}\log\xi. 
\end{equation}
Since $\Pi^{(1)}$ is a solution of (\ref{PF-WL-fCS_2}), we find 
\begin{equation}
\Bigl[ \xi\partial_\xi-\xi A \Bigr]\left( \Pi^{(1)}\log\xi \right)\ =\ \Pi^{(1)}. 
\end{equation}
Then, by choosing $\Pi^{(2)}_0=v_2$, the PF equation reduces to 
\begin{equation}
\Bigl[ nI-A_0 \Bigr]\Pi^{(2)}_n+c\Pi^{(1)}_{n-1}\ =\ \sum_{m=0}^{n-1}A_{n-m}\Pi^{(2)}_m, 
\end{equation}
for $n\ge1$. 
Now, the condition for $n=1$ can be solved by 
\begin{equation}
c\ =\ \frac2{\nu+4}, \hspace{1cm} \Pi^{(2)}_1\ =\ \frac1{4(\nu+4)}
\left[
\begin{array}{c}
\nu-20 \\
2(5\nu-4) \\
-4(\nu^2-\nu+4)
\end{array}
\right]. 
\end{equation}
The vectors $\Pi^{(2)}_n$ for $n\ge2$ are then determined recursively. 
Note that we could have added to $\Pi^{(2)}_1$ any vector proportional to $v_1$ since it is annihilated by $I-A_0$. 
This ambiguity just corresponds to adding a multiple of $\Pi^{(1)}$ to $\Pi^{(2)}$. 

There is one more solution 
\begin{equation}
\Pi^{(3)}\ =\ \sum_{m=0}^\infty \Pi^{(3)}_n\xi^n+\frac1{\nu+4}\Pi^{(1)}\log\xi, \hspace{1cm} \Pi^{(3)}_0\ =\ v_3\ =\ 
\left[
\begin{array}{c}
0 \\
1 \\
-\frac{\nu+4}3
\end{array}
\right], 
\end{equation}
which can be determined similarly. 

The general solution of (\ref{PF-WL-fCS_2}) is a linear combination of $\Pi^{(1)},\Pi^{(2)},\Pi^{(3)}$, from which we find the physically meaningful solution $\Pi$. 
As mentioned at the beginning, we require that $\Pi$ vanishes at $\xi=0$. 
This implies 
\begin{eqnarray}
\Pi 
&\propto& \Pi^{(1)} \nonumber \\
&=& \frac1{\nu+4}
\left[
\begin{array}{l}
(\nu+4)\xi+\xi^2 \\ [1mm] 
(2\nu+8)\xi+(3\nu+14)\xi^2+\frac12(\nu+12)\xi^3+\frac34\xi^4  \\ [1mm] 
(4\nu+16)\xi+(12\nu+52)\xi^2+(12\nu+64)\xi^3+\frac14(15\nu+132)\xi^4
\end{array}
\right]+{\cal O}(\xi^5). \nonumber \\
\end{eqnarray}
Indeed, this coincides with the exact result (\ref{exact WL-fCS}) up to an overall constant. 
From this expression, we find $\delta_3=1$ and $\delta_4=0$. 
Note that $\Pi^{(1)}$ does not give the leading term in the general solution. 
These exponents are therefore selected by a physical input. 

The coefficient in front of $\Pi^{(1)}$ cannot be determined by (\ref{PF-WL-fCS_2}). 
However, it is not just a number but a function of the parameter $\nu$ in general, so this ambiguity is quite large. 
As mentioned at the end of section \ref{PF eqs}, this ambiguity is almost fixed by another equation 
\begin{equation}
\partial_\nu\Pi\ =\ A_\nu\Pi, 
\end{equation}
where $A_\nu$ is given in (\ref{nu-PF-WL}). 
This must be satisfied by $\Pi=c_1(\nu)\Pi^{(1)}$ for a suitable choice of $c_1(\nu)$. 
We find that $c_1(\nu)$ must satisfy 
\begin{equation}
(\nu+4)\partial_\nu c_1(\nu)\ =\ c_1(\nu). 
\end{equation}
The general solution of this equation is $c_1(\nu)=c_1(\nu+4)$. 
This gives, for example 
\begin{equation}
\pi_0\ =\ c_1\Bigl[ (\nu+4)\xi+\xi^2 \Bigr]+{\cal O}(\xi^5). 
\end{equation}

To fix the remaining constant $c_1$, recall that $\nu$ is proportional to the number $N_f$ of flavors. 
Therefore, $\nu=0$ corresponds to CS theory for which $\pi_0$ is given as \cite{Suyama:2025ydf} 
\begin{equation}
\pi_0\ =\ -(a+a^{-1}-2). 
\end{equation}
By using $\gamma=a^{1/2}+a^{-1/2}$, we find $c_1=-1$. 

\vspace{5mm}

Next, we determine the local solution for $T$. 
Since the PF equation for 't~Hooft coupling (\ref{PF-tH}) is an inhomogeneous equation, we need to determine a particular solution of this equation and the general solution for the homogeneous equation 
\begin{equation}
\partial_\xi T\ =\ \tilde AT. 
\end{equation}

This homogeneous equation can be solved in the same manner as for $\Pi$. 
Recall that the residue matrix $\tilde A_0$ of $\tilde A$ is diagonalizable with the eigenvalues $\frac12,0,0$. 
Let $\tilde v_1,\tilde v_2, \tilde v_3$ be the corresponding eigenvectors, respectively. 
There is no resonance, so the solutions are of the form 
\begin{equation}
T^{(1)}\ =\ \sum_{n=0}^\infty T^{(1)}_n\xi^{n+1/2}, \hspace{1cm} T^{(2)}\ =\ \sum_{n=0}^\infty T^{(2)}_n\xi^n, \hspace{1cm} T^{(3)}\ =\ \sum_{n=0}^\infty T^{(3)}_n\xi^n, 
\end{equation}
with $T^{(i)}_0=\tilde v_i$. 

Using $\Pi$ obtained above, the inhomogeneous equation (\ref{PF-tH}) can be written as 
\begin{equation}
\Bigl[ \xi\partial_\xi-\xi\tilde A \Bigr]T_0\ =\ \xi J\Pi\ =\ \frac{\nu+4}8\tilde v_1\xi+{\cal O}(\xi^2). 
\end{equation}
Since 
\begin{equation}
\Bigl[ \xi\partial_\xi-\xi\tilde A \Bigr]\tilde v_i\xi^\rho\ =\ (\rho-\tilde\lambda_i)\tilde v_i\xi^\rho+{\cal O}(\xi^{\rho+1}) 
   \label{tool-ih}
\end{equation}
hold, where $\tilde\lambda_i$ is the eigenvalue of $\tilde A_0$ for $\tilde v_i$, we find that 
\begin{equation}
T_0\ =\ \frac{\nu+4}4\tilde v_1\xi+{\cal O}(\xi^2)
\end{equation}
is a solution. 
Higher-order terms in $T_0$ are also determined similarly. 

The physically relevant solution $T$ is a linear combination of them. 
We require that $T$ vanishes at $\xi=0$, as for $\Pi$. 
This implies that $T^{(2)}$ and $T^{(3)}$ must be excluded from $T$. 
In addition, we know from \cite{Suyama:2025ydf} that $t$ does not contain terms which are non-analytic at $\xi=0$ when we set $\nu=0$. 
This also excludes $T^{(1)}$. 
Therefore, we obtain 
\begin{equation}
T\ =\ T_0\ =\ 
\left[
\begin{array}{l}
\frac{1}4(\nu+4)\xi-\frac{1}8(\nu+2)\xi^2+\frac{1}{48}(3\nu+4)\xi^3 \\ [2mm] 
\frac{1}2(\nu+4)\xi+\frac{1}2(\nu+5)\xi^2-\frac16\nu\xi^3 \\ [2mm] 
(\nu+2)\xi+\frac{1}2(5\nu+22)\xi^2+\frac{1}{12}(19\nu+116)\xi^3
\end{array}
\right]
+{\cal O}(\xi^4). 
\end{equation}
This coincides with the exact result (\ref{exact tH-fCS}), as expected. 

We find from this result that the exponents take $\delta_1=1$ and $\delta_2=0$. 
Again, this is selected by a physical input, since $T_0$ does not give the leading contribution in the general solution of (\ref{PF-tH}). 
The values of $\delta_a$ obtained so far imply $W_1={\cal O}(t^0)$. 
This is of course expected since the perturbative calculation gives $W_1=1+{\cal O}(t)$. 

\vspace{5mm}

\subsection{fCS theory at $\gamma=0$}   \label{local-inf-fCS}

\vspace{5mm}

The local solution for fCS theory at $\gamma=0$ can be obtained similarly, after the gauge transformation (\ref{gauge-fCS}). 
However, unlike the case $\gamma=2$, we do not have any physical input for $\Pi$ and $T$ at $\gamma=0$. 
Therefore, we cannot fix the integration constants completely unless we compare the local solution with the exact results (\ref{exact WL-fCS})(\ref{exact tH-fCS}). 
Nevertheless, we will see below that we can guess the qualitative behavior of the solution in the limit $\gamma\to0$, assuming that the physical solution is ``generic'' at $\gamma=0$. 

\vspace{5mm}

First, consider the PF equation (\ref{rescaled PF-WL}) for $\hat\Pi$. 
We choose a local coordinate $\xi=\gamma$. 
The residue matrix ${\cal A}_0$ of $\cal A$ is diagonalizable with the eigenvalues $-\frac12,-1,-3$. 
The corresponding solutions are of the forms 
\begin{eqnarray}
\hat\Pi^{(1)} 
&=& \sum_{n=0}^\infty \Pi^{(1)}_n\xi^{n-1/2}, \\
\hat\Pi^{(2)} 
&=& 
\left[
\begin{array}{c}
\xi^{-1} \\
0 \\
0
\end{array}
\right], \\
\hat\Pi^{(3)} 
&=& \sum_{n=0}^\infty \Pi^{(3)}_n\xi^{n-3}, 
\end{eqnarray}
respectively. 
Remarkably, $\hat\Pi^{(2)}$ is an exact solution. 
Note that $\hat\Pi^{(3)}$ could have $\log\xi$ factor due to the resonance, but it turns out to be absent. 

The physical solution $\hat \Pi$ is a linear combination of them. 
We substitute $\hat \Pi=c_1(\nu)\hat \Pi^{(1)}+c_2(\nu)\hat\Pi^{(2)}+c_3(\nu)\hat\Pi^{(3)}$ into the equation (\ref{rescaled PF-WL}) with $\partial_\gamma$ replaced with $\partial_\nu$. 
In order for this $\hat\Pi$ to be a solution, the following conditions 
\begin{eqnarray}
2\nu\partial_\nu c_1(\nu) 
&=& c_1(\nu), \\ 
2\nu\partial_\nu c_2(\nu)+3\nu\partial_\nu c_3(\nu) 
&=& 2c_2(\nu)+2c_3(\nu), \\
\partial_\nu c_3(\nu) 
&=& 0, 
\end{eqnarray}
must be satisfied. 
The general solution of these equations is 
\begin{equation}
c_1(\nu)\ =\ c_1\nu^{1/2}, \hspace{1cm} 
c_2(\nu)\ =\ c_2\nu-c_3, \hspace{1cm} 
c_3(\nu)\ =\ c_3. 
\end{equation}
Then, the leading term of $\hat\Pi$ is found to be 
\begin{equation}
\hat\Pi\ \sim\ 
\left[
\begin{array}{c}
\frac12(2c_2\nu+c_3)\xi^{-1} \\ [1mm] 
\frac12c_3\xi^{-2} \\ [1mm] 
c_3\xi^{-3}
\end{array}
\right]. 
\end{equation}
By the gauge transformation $\Pi=U\hat\Pi$, we find that $\pi_0$ behaves as 
\begin{equation}
\pi_0\ \sim\ \frac12(2c_2\nu+c_3). 
\end{equation}
Therefore, the exponents take $\delta_3=\delta_4=0$, implying $W_1={\cal O}(t^{-1})$. 

If we compare $\hat\Pi$ with the exact result (\ref{exact WL-fCS}), then we find $c_1=0$, $c_2=2$, $c_3=8$. 
For these values, the coefficients in the leading term of $\hat\Pi$ do not vanish. 
Therefore, the leading term of the general solution gives the leading term of the physical solution in this case. 

We expect that this occurs in general, except for cases in which a physical requirement makes the leading term to vanish. 
In the previous subsection, we selected such a special solution at $\gamma=2$. 
The local solutions around the other singularities can be obtained from this special solution at $\gamma=2$ via the analytic continuation. 
This amounts to employing the connection coefficients which are typically complicated functions of $\nu$. 
Then, the analytically continued solution would not be special, so that the leading term is given by the one of the general local solution, as we have observed at $\gamma=0$. 
If this is really the general case, then we can guess the qualitative behavior of the physical solution at each singularity without the knowledge of the connection coefficients. 
Of course, the complete knowledge of local solutions for the undeformed theory, if it is available, allows us to fix the integration constants. 

\vspace{5mm}

Next, we solve the PF equation (\ref{rescaled PF-tH}) for $\hat T$. 
The eigenvalues of the residue matrix $\tilde {\cal A}_0$ of $\tilde{\cal A}$ are $-1,-2,-3$, and the corresponding solutions of the homogeneous equation are 
\begin{eqnarray}
\hat T^{(1)} 
&=& 
\left[
\begin{array}{l}
\xi^{-1} \\ [1mm] 
\frac12\nu\xi^{-1}+\frac12 \\ [1mm] 
\frac14\nu+\frac38\xi
\end{array}
\right], \\
\hat T^{(2)} 
&=& \sum_{n=0}^\infty \hat T^{(2)}_n\xi^{n-2}-\frac1\nu \hat T^{(1)}\log\xi, \\
\hat T^{(3)} 
&=& \sum_{n=0}^\infty \hat T^{(3)}_n\xi^{n-3}-\frac{\nu^2+8}{8\nu^2}\hat T^{(1)}\log\xi, 
\end{eqnarray}
where $\hat T^{(1)}$ is an exact solution. 
Note that there could have been a term in $\hat T^{(3)}$ which is proportional to $\hat T^{(2)}\log\xi$, but it turns out to be absent. 

In order to obtain a particular solution $\hat T_0$ of (\ref{rescaled PF-tH}), we rewrite the inhomogeneous equation as 
\begin{eqnarray}
\Bigl[ \xi\partial_\xi-\xi\tilde {\cal A} \Bigr]\hat T_0\ =\ \xi \hat J \hat\Pi
&=& \sum_{n=0}^\infty(c_{1n}\tilde v_1+c_{2n}\tilde v_2+c_{3n}\tilde v_3)\xi^{n-2} \nonumber \\
& & +\sum_{n=0}^\infty(c'_{1n}\tilde v_1+c'_{2n}\tilde v_2+c'_{3n}\tilde v_3)\xi^{n-1/2}, 
\end{eqnarray}
where $\tilde v_i$ are the eigenvectors of $\tilde {\cal A}_0$. 
We can determine $\hat T_0$ by using equations similar to (\ref{tool-ih}). 
Namely, if there is a term proportional to $\tilde v_i\xi^\rho$ in the right-hand side, then we add the same term to $\hat T_0$ with a suitable coefficient. 
In fact, this is not enough for determining $\hat T_0$. 
For example, a term $\tilde v_1\xi^{-1}$ satisfies 
\begin{equation}
\Bigl[ \xi\partial_\xi-\xi\tilde{\cal A} \Bigr]\tilde v_1\xi^{-1}\ =\ {\cal O}(\xi^0), 
\end{equation}
since the eigenvalue of $\tilde v_1$ is $-1$. 
Since this term does appear in the right-hand side, instead of (\ref{tool-ih}), we need to use 
\begin{equation}
\Bigl[ \xi\partial_\xi-\xi\tilde {\cal A} \Bigr]\hat T^{(1)}\log\xi\ =\ \hat T^{(1)}\ =\ \tilde v_1\xi^{-1}+{\cal O}(\xi^0), 
   \label{extra log}
\end{equation}
which introduces a $\log\xi$ factor to $\hat T_0$. 
Then, we find 
\begin{equation}
\hat T_0\ =\ 
\left[
\begin{array}{l}
\displaystyle -\frac{c_3}{8\nu\xi}+\frac{c_3(5\nu^2+8)}{64\nu^2}\log\xi+\frac{8c_2\nu^3+c_3(5\nu^2-8)}{32\nu^2} \\ [4mm] 
\displaystyle \frac{c_3}{4\nu\xi}-\frac{c_3(5\nu^2+8)}{32\nu^2}\log\xi-\frac{4c_2\nu^3+c_3(3\nu^2-4)}{8\nu^2} \\ [4mm] 
\displaystyle -\frac{c_3}{2\nu\xi}+\frac{c_3(5\nu^2+8)}{16\nu^2}\log\xi+\frac{8c_2\nu^3+c_3(7\nu^2-8)}{8\nu^2}
\end{array}
\right]
+{\cal O}(\xi^{1/2}). 
\end{equation}

The physical solution is of the form $\hat T=\tilde c_1(\nu)\hat T^{(1)}+\tilde c_2(\nu)\hat T^{(2)}+\tilde c_3(\nu)\hat T^{(3)}+\hat T_0$. 
The conditions for these coefficients are 
\begin{eqnarray}
64\nu^3\partial_\nu\tilde c_1(\nu) 
&=& 64\nu\tilde c_2(\nu)+(-8\nu^2+64)\tilde c_3(\nu)+c_3(5\nu^2-8), \\ [1mm] 
4\nu^2\partial_\nu\tilde c_2(\nu)+4\nu\partial_\nu\tilde c_3(\nu) 
&=& 4\nu\tilde c_2(\nu)+8\tilde c_3(\nu)-c_3, \\ [1mm]  
\partial_\nu\tilde c_3(\nu) 
&=& 0. 
\end{eqnarray}
The general solution is 
\begin{eqnarray}
\tilde c_1(\nu) 
&=& \frac{64\tilde c_2-8\tilde c_3+5c_3}{64}\log\nu+\tilde c_1, \\
\tilde c_2(\nu) 
&=& \tilde c_2\nu-\frac{8\tilde c_3-c_3}{8\nu}, \\ 
\tilde c_3(\nu) 
&=& \tilde c_3. 
\end{eqnarray}

Finally, we find 
\begin{equation}
T\ =\ U\hat T\ =\ 
\left[
\begin{array}{l}
\displaystyle \frac{\tilde c_2\nu}{\xi}+\frac{5c_3-64\tilde c_2-8\tilde c_3}{64}\log\xi \\ [4mm] 
\displaystyle -\frac{2\tilde c_2\nu}{\xi}-\frac{5c_3-64\tilde c_2-8\tilde c_3}{32}\log\xi \\ [4mm] 
\displaystyle \frac{4\tilde c_2\nu}\xi+\frac{5c_3-64\tilde c_2-8\tilde c_3}{16}\log\xi
\end{array}
\right]
+{\cal O}(\xi^0). 
\end{equation}
The top entry of $T$ gives $t$ which coincides with the exact result (\ref{exact tH-fCS}) if we choose 
\begin{equation}
\tilde c_1\ =\ -2\log2, \hspace{1cm} \tilde c_2\ =\ -1, \hspace{1cm} \tilde c_3\ =\ -3. 
\end{equation}
Note that the physical solution for $T$ at $\gamma=0$ has the leading term which is also dominant in the general solution, as expected. 
The exponents are found to be $\delta_1=-1$ and $\delta_2=0$. 
Unfortunately, the leading term of $T$ vanishes at $\nu=0$, so that we cannot fix $\tilde c_2$ by the result for CS theory. 

\vspace{5mm}

The results so far indicate that the singularity at $\gamma=0$ corresponds to a strong coupling limit $t\to\infty$ at which the Wilson loop vanishes. 
This is quite different from the behavior exhibited by a theory with a gravity dual. 

\vspace{5mm}

\subsection{ABJ theory at $\gamma=2-2e^B$}

\vspace{5mm}

Now we turn to ABJ theory. 
Recall that ABJ theory depends on two 't~Hooft couplings $t_1$ and $t_2$, and has two sets of Wilson loops $W^{(1)}_n$ and $W^{(2)}_n$. 
At the point $\gamma=2-2e^B$, it turns out that the integrals (\ref{integral-WL}) and (\ref{integral-tH}) for $t_1W^{(1)}_n$ and $t_1$, respectively, vanish since the integrands inside the integration contour $C$ become analytic. 
On the other hand, $t_2W^{(2)}_n$ and $t_2$ vanish at the other singularity $\gamma=-2+2e^B$. 

In fact, there are no new ingredients in obtaining the local solutions around these singularities. 
Nevertheless, we will see that the Wilson loops and the 't~Hooft couplings have interesting dependence on the parameter $B$ which cannot be obtained by perturbative methods. 

\vspace{5mm}

Let $p=2-2e^B$. 
We take a local coordinate $\xi=\gamma-p$. 
The residue matrix $A_0$ has eigenvalues $1,0,0$. 
The corresponding solutions have of the forms 
\begin{eqnarray}
\Pi^{(1)} 
&=& \sum_{n=0}^\infty \Pi^{(1)}_n\xi^{n+1}, \\
\Pi^{(2)} 
&=& \sum_{n=0}^\infty \Pi^{(2)}_n\xi^{n}+\frac{3-e^{-B}}{16}\Pi^{(1)}\log\xi, \\
\Pi^{(3)} 
&=& \sum_{n=0}^\infty \Pi^{(3)}_n\xi^{n}+\frac1{16}e^{-B}\Pi^{(1)}\log\xi. 
\end{eqnarray}

It turns out that the conditions for a linear combination $\Pi=c_1(B)\Pi^{(1)}+c_2(B)\Pi^{(2)}+c_3(B)\Pi^{(3)}$ to have the correct $B$-dependence are rather complicated. 
At the leading order in $\xi$, we have 
\begin{eqnarray}
2(1-e^B)\partial_Bc_2(B) 
&=& -(1-e^B)c_2(B)+c_3(B), \\ [1mm] 
(1-e^B)\partial_Bc_3(B) 
&=& -(1-e^B)c_2(B)+(1-2e^B)c_3(B). 
\end{eqnarray}
We introduce $c_0(B)$ satisfying 
\begin{equation}
2(1-e^B)\partial_B^2c_0(B)-(1+3e^B)\partial_Bc_0(B)\ =\ 0, 
\end{equation}
by which $c_2(B)$ and $c_3(B)$ can be written as 
\begin{eqnarray}
c_2(B) 
&=& (1-e^B)c_0(B), \\ [1mm] 
c_3(B) 
&=& 2(1-e^B)^2\partial_Bc_0(B)+(1-e^B)(1-3e^B)c_0(B). 
\end{eqnarray}
The general solution for $c_0(B)$ is 
\begin{equation}
c_0(B)\ =\ c_2+c_3\left[ \frac12\log\frac{1+e^{B/2}}{1-e^{B/2}}-\frac1{2\sinh(B/2)} \right]. 
\end{equation}
This is a bit complicated. 
In order to simplify the following calculations, we use the knowledge that $B=0$ corresponds to ABJM theory for which both $t_1W^{(1)}_n$ and $t_2W^{(2)}$ vanish in the limit $\xi\to0$. 
This is the case only when we set $c_3=0$. 
Then, we obtain 
\begin{equation}
c_2(B)\ =\ c_2(1-e^B), \hspace{1cm} c_3(B)\ =\ c_2(1-e^B)(1-3e^B). 
\end{equation}

At the next order in $\xi$, we find the last condition 
\begin{equation}
32e^B\partial_Bc_1(B)\ =\ -16e^Bc_1(B)+c_2(21e^{2B}-6e^B-3) 
\end{equation}
under the above choice of $c_2(B)$ and $c_3(B)$. 
The general solution is 
\begin{equation}
c_1(B)\ =\ c_1e^{-B/2}+c_2\left[ \frac7{16}e^B-\frac38+\frac3{16}e^{-B} \right]. 
\end{equation}

The integration constants $c_1$ and $c_2$ can be different between $t_1W^{(1)}_n$ and $t_2W^{(2)}_n$. 
Since $t_1W^{(1)}_n$ vanishes at $\xi=0$, we need to set $c_2=0$. 
For this case, we obtain 
\begin{equation}
\Pi\ =\ c_1e^{-B/2}
\left[
\begin{array}{c}
1 \\
2 \\
4
\end{array}
\right]
\xi+{\cal O}(\xi^2). 
   \label{local-ABJ-WL}
\end{equation}
By the comparison with the result for ABJM theory in \cite{Suyama:2025ydf}, we find $c_1=-1$. 
The exponents for this solution are $\delta_3=1$ and $\delta_4=0$. 
On the other hand, $t_2W^{(2)}_n$ does not need to vanish at $\xi=0$. 
Therefore, we obtain 
\begin{equation}
\Pi\ =\ c_2(1-e^B)
\left[
\begin{array}{c}
1 \\
1-3e^B \\
2(1-4e^B+5e^{2B})
\end{array}
\right]
+{\cal O}(\xi). 
\end{equation}
We expect that $c_2$ is non-vanishing. 
Then, we find $\delta_3=\delta_4=0$. 
Since $t_2$ must be non-zero at $\xi=0$ for a non-zero $B$, the above result of $W^{(2)}_1$ is non-perturbative in terms of $t_2$. 

\vspace{5mm}

The local solution for the 't~Hooft couplings can be obtained similarly. 
We require the analyticity of $T$ at $\xi=0$ since this is the case for $B=0$. 
Then we obtain 
\begin{equation}
\tau_0\ =\ \frac{c_2}8-\frac{c_2}4B+\tilde c_3-\tilde c_2\log\frac{\sqrt{1-e^B}+1}{\sqrt{1-e^B}-1}+{\cal O}(\xi). 
\end{equation}
We notice that $B$, not only $e^B$, appears in $\tau_0$ due to the integrations with respect to $B$ which are needed in determining the $B$-dependence of $T$. 

The integration constants $c_2,\tilde c_2,\tilde c_3$ are different between $t_1$ and $t_2$ since they corresponds to different choices of the integration contour in (\ref{integral-WL})(\ref{integral-tH}). 
The above expression for $\tau_0$ allows us to choose the constants such that $t_1$ vanishes at $\xi=0$ while $t_1-t_2=B$ holds. 
In particular, we find $c_2=-4$. 
The exponents for $t_1$ turn out to be $\delta_1=1$ and $t_2=0$, while for $t_2$ we find $\delta_1=\delta_2=0$. 
This implies that both $W^{(1)}_n$ and $W^{(2)}_n$ are finite at $\xi=0$. 

\vspace{5mm}

\subsection{ABJ theory at $\gamma=\infty$}

\vspace{5mm}

In this case, we observed in section \ref{residue matrix} that the residue matrices ${\cal A}_0$ and $\tilde {\cal A}_0$ have non-trivial Jordan blocks. 
This turns out to be another source of $\log\xi$ factor in the local solutions, as we will see below. 

\vspace{5mm}

Recall that there is a basis consisting of $v_1=v(0),v_2=v(1),v_3$ on which ${\cal A}_0$ acts as 
\begin{equation}
{\cal A}_0v_1\ =\ 2v_1, \hspace{1cm} {\cal A}_0v_2\ =\ 0, \hspace{1cm} {\cal A}_0v_3\ =\ v_2, \hspace{1cm} v_3\ =\ 
\left[
\begin{array}{c}
-\frac12 \\
0 \\
\frac13
\end{array}
\right]
.
\end{equation}
Since $v_1$ and $v_2$ are eigenvectors of ${\cal A}_0$, the corresponding solutions can be obtained as before. 
By choosing $\xi=\gamma^{-1}$ as a local coordinate, they are of the forms 
\begin{equation}
\hat \Pi^{(1)}\ =\ \sum_{n=0}^\infty \hat \Pi^{(1)}_n\xi^{n+2}, \hspace{1cm} \hat \Pi^{(2)}\ =\ \sum_{n=0}^\infty \hat \Pi^{(2)}_n\xi^n. 
\end{equation}
On the other hand, since $v_3$ is not an eigenvector of ${\cal A}_0$, the ansatz for $\hat\Pi^{(3)}$ must be further modified, in addition to taking the resonance into account. 
We find that $\hat \Pi^{(3)}$ is of the form 
\begin{equation}
\hat \Pi^{(3)}\ =\ \sum_{n=0}^\infty\hat\Pi^{(3)}_n\xi^n+\hat\Pi^{(2)}\log\xi+2(2e^{2B}-1)\hat\Pi^{(1)}\log\xi. 
   \label{Pi^3-ABJ}
\end{equation}
Due to the presence of the second term in the right-hand side, the PF equation (\ref{rescaled PF-WL}) at order ${\cal O}(\xi^0)$ becomes 
\begin{equation}
\hat\Pi^{(2)}_0\ =\ {\cal A}_0\hat\Pi^{(3)}_0. 
\end{equation}
This is satisfied by the choice $\hat\Pi^{(3)}_0=v_3$. 
Then $\hat\Pi^{(3)}_n$ with $n\ge1$ are determined recursively. 
The last term in the right-hand side of (\ref{Pi^3-ABJ}) is due to the resonance. 
Note that the $\log\xi$ factor appears in the leading term of $\hat\Pi^{(3)}$ which does not occur when the residue matrix is diagonalizable. 

The solution with the correct $B$-dependence turns out to be 
\begin{equation}
\hat\Pi\ =\ \Bigl[ c_1+\left( 4c_2+2c_3(2B-1) \right)e^{2B} \Bigr]\hat\Pi^{(1)}+(c_2+c_3B)\hat\Pi^{(2)}+c_3\hat\Pi^{(3)}. 
\end{equation}
The Wilson loops are obtained by $\Pi=U\hat\Pi$ where the gauge transformation $U$ is given as (\ref{gauge-ABJ}). 
For example, we find 
\begin{equation}
\pi_0\ =\ \left[ \frac12(2c_2+c_3(2B-1)+c_3\log\xi \right]\xi^{-1}+{\cal O}(\xi\log\xi). 
   \label{pi_0-ABJ}
\end{equation}
The exponents for this solution are $\delta_3=-1$ and $\delta_4=1$. 
This is the first case with $\delta_4\ne0$ in our analysis. 

Next, we consider 't~Hooft couplings. 
The residue matrix $\tilde {\cal A}_0$ again has a non-trivial Jordan block. 
The homogeneous solutions are 
\begin{eqnarray}
\hat T^{(1)} 
&=& \sum_{n=0}^\infty \hat T^{(1)}_n\xi^{n+2}, \\ [1mm] 
\hat T^{(2)} 
&=& \sum_{n=0}^\infty \hat T^{(2)}_n\xi^{n+1}, \\ [1mm] 
\hat T^{(3)} 
&=& \sum_{n=0}^\infty \hat T^{(3)}_n\xi^{n+1}+\hat T^{(2)}\log\xi. 
\end{eqnarray}
Note that there are no terms due to the resonance. 
Since the inhomogeneous term ${\cal J}\hat\Pi$ now contains terms with $\log\xi$ factor, the determination of a particular solution $\hat T_0$ is a bit more complicated. 
We find 
\begin{equation}
\hat T_0\ =\ -\frac{c_3}4\tilde v_1\xi+\frac12(c_2+c_3(B-1))\hat T^{(2)}\log\xi+\frac{c_3}2\hat T^{(3)}\log\xi+{\cal O}(\xi^3\log\xi). 
\end{equation}
Note that $(\log\xi)^2$ factor appears from $\hat T^{(3)}\log\xi$, which is absent in the previous cases. 

The solution with the correct $B$-dependence turns out to be 
\begin{equation}
\hat T\ =\ \tilde c_1\hat T^{(1)}+\left[ \tilde c_2+\frac14(4\tilde c_3-c_3)B \right]\hat T^{(2)}+\tilde c_3\hat T^{(3)}+\hat T_0
\end{equation}
By the gauge transformation $T=U\hat T$, this gives 
\begin{equation}
\tau_0\ =\ \frac{c_3}2(\log\xi)^2+\frac12\Bigl[ c_2+c_3(B-2)+2\tilde c_3 \Bigr]\log\xi+{\cal O}(\xi^0). 
   \label{tau_0-ABJ}
\end{equation}
This indicates that both $t_1$ and $t_2$ diverge in the limit $\xi\to0$. 
This is compatible with the relation $t_1-t_2=B$. 
Note that the leading term proportional to $(\log\xi)^2$ comes from $\hat T_0$. 
The comparison with ABJM theory \cite{Suyama:2025ydf} implies $c_3=2i/\pi$. 

The exponents for this solution are $\delta_1=0$ and $\delta_2=2$. 
The vanishing of $\delta_1$ ensures that the Wilson loop grows exponentially, and the value of $\delta_2$ gives the correct dependence on the 't~Hooft couplings compatible with the presence of the semi-classical gravity dual \cite{Marino:2009jd}. 
Indeed, for $B=0$ we obtain 
\begin{equation}
W_1\ \sim\ \exp\left( \pi\sqrt{2\lambda} \right). 
\end{equation}
Note that we chose the sign of $(t/c_t)^{1/\delta_2}$ such that $W_1$ diverges in the strong coupling limit, as the local solutions (\ref{pi_0-ABJ})(\ref{tau_0-ABJ}) indicate. 

\vspace{1cm}

\section{Physical path}   \label{para dependence}

\vspace{5mm}

So far, we have treated $\gamma$ as a complex variable. 
However, most of the values of $\gamma$ do not make physical sense. 
Recall the definition (\ref{parameters-fCS}) of the parameters in fCS theory. 
This shows that the physically relevant values of $\gamma$ must satisfy ${\rm Re}(t)=0$. 
This condition defines a path which starts from the point $\gamma=2$ corresponding to the weak coupling limit in the $\gamma$-plane. 
We call this the physical path. 
The shape of this path depends on $\nu$ whose physical value is purely imaginary. 
For ABJ theory, we should require ${\rm Re}(t_1)={\rm Re}(t_2)=0$. 
Since we have the relation $t_1-t_2=B$, as long as $B$ is chosen to be purely imaginary, we only need to require, say, ${\rm Re}(t_1)=0$. 
This defines the physical path for ABJ theory. 

In the previous section, we obtained local solutions around the regular singularities of the PF equations. 
Each of them has some physical meaning only when the corresponding singularity lies on the physical path.  In this section, we investigate the physical paths for fCS and ABJ theories to understand their parameter dependence and also the physical meanings of the singularities. 

\vspace{5mm}

First, we consider fCS theory. 
As explained in appendix \ref{app-exact}, we have an exact formula  
\begin{equation}
t\ =\ 2\log\frac\gamma2+\frac\nu2\left( 1-\frac2\gamma \right). 
   \label{t-fCS}
\end{equation}
It is easy to notice that $t$ vanishes at $\gamma=2$ which corresponds to the weak coupling limit, as mentioned in section \ref{algebraic curve}. 

\begin{figure}
\begin{center}
\includegraphics[width=70mm]{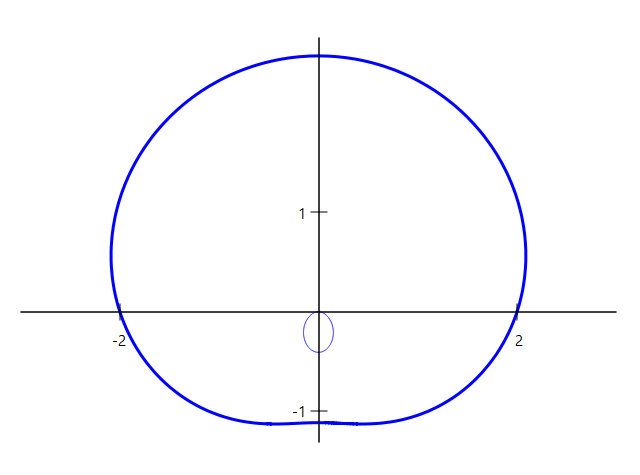}
\includegraphics[width=70mm]{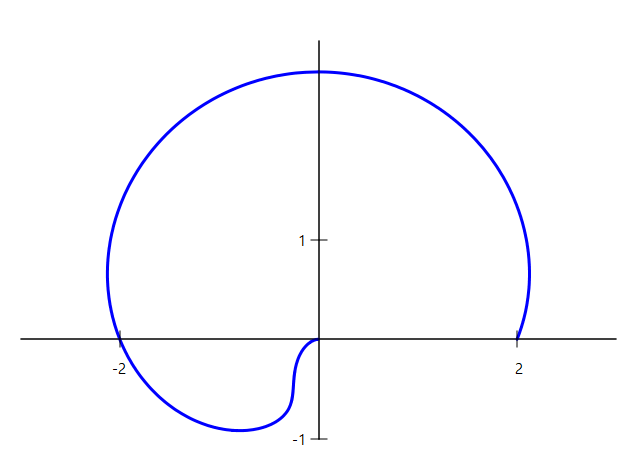}
\end{center}
\caption{
The physical paths for fCS theory. 
The left figure is the path for $\nu_0=1.3$. 
There also exists a disconnected path starting from and arriving at $\gamma=0$. 
The right figure is the path for $\nu_0=1.6$. 
}
   \label{physical path}
\end{figure}

\begin{figure}
\begin{center}
\includegraphics[width=70mm]{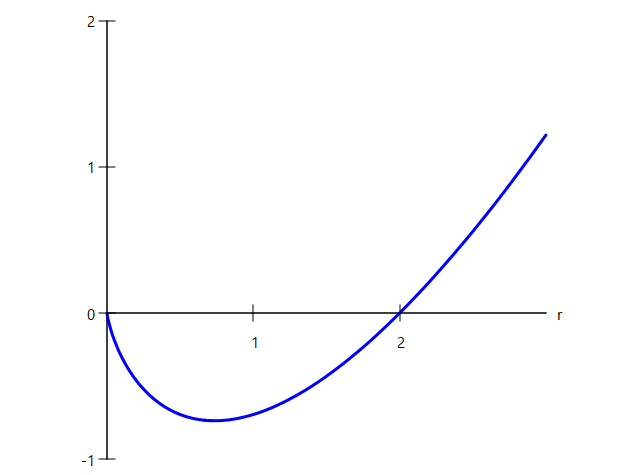}
\end{center}
\caption{
The graph of $r\log\frac r2$. 
The minimum value $-2e^{-1}$ is attained at $r=2e^{-1}$. 
}
   \label{graph}
\end{figure}

We investigate the physical path by using (\ref{t-fCS}). 
In the following, we assume $\nu=\nu_0i$ with $\nu_0\ge0$. 
By using the polar coordinate $\gamma=re^{i\theta}$, the condition ${\rm Re}(t)=0$ for the physical path can be written as 
\begin{equation}
r\log\frac r2\ =\ \frac{\nu_0}2\sin\theta. 
   \label{phys-curve-fCS}
\end{equation}
The physical path is depicted in Figure \ref{physical path}. 
The graph of $r\log\frac r2$ given in Figure \ref{graph} is helpful for understanding the shape of the physical path. 

The physical path starts from $\gamma=2$ and arrives at $\gamma=-2$ through the upper-half $\gamma$-plane. 
At $\gamma=-2$, we find 
\begin{equation}
t\ =\ 2\pi i+\nu. 
   \label{boundary-fCS}
\end{equation}
As reviewed in the introduction, it is known \cite{Giveon:2008zn}\cite{Suyama:2011yz} that the supersymmetry is preserved as long as 
\begin{equation}
N\ \le\ k+N_f
   \label{SUSY-fCS}
\end{equation}
is satisfied, assuming that the Chern-Simons level $k$ is positive. 
This shows that the singularity at $\gamma=-2$ corresponds to the point where the supersymmetry is broken as we increase ${\rm Im}(t)$. 

We also have an exact formula 
\begin{equation}
\pi_0\ =\ -(\gamma-2)(\gamma+\nu+2)
\end{equation}
given in appendix \ref{app-exact}. 
We find $\pi_0=4\nu$ at $\gamma=-2$, which implies 
\begin{equation}
W_1\ =\ -\frac\nu{\nu+2\pi i}. 
\end{equation}
Note that this is a real value for a physical $\nu$. 
This vanishes for $\nu=0$, indicating the confinement in CS theory discussed in \cite{Witten:1999ds}. 
The non-vanishing of $W_1$ for $\nu\ne0$ is then naturally understood as the screening of the Wilson loop by the matter in the fundamental representation. 

The physical path can be extended beyond $\gamma=-2$. 
It is curious to notice that the shape of the physical path depends on whether $\nu_0$ is smaller or larger than $4e^{-1}$. 
\begin{itemize}
\item 
For $\nu_0<4e^{-1}$, the equation (\ref{phys-curve-fCS}) has a solution for any values of $\theta$. 
Then, the physical path starting from $\gamma=2$ forms a closed curve, as in the left of Figure \ref{physical path}. 
The endpoint corresponds to $\theta=2\pi$ at which we have $t=4\pi i$. 
Therefore, the strong coupling limit does not correspond to any specific point in the $\gamma$-plane, but to going along this closed curve infinitely many times. 
This is similar to CS theory. 
Since $|\pi_0|$ is bounded on the physical path, $W_1$ behaves as ${\cal O}(t^{-1})$ in the strong coupling limit. 
\item 
For $\nu_0>4e^{-1}$, on the other hand, the allowed values of $\theta$ are limited since the minimum value taken by the right-hand side in (\ref{phys-curve-fCS}) is smaller than that of the left-hand side. 
Let $\theta_*>\pi$ be the smallest positive value of $\theta$ for which the equation (\ref{phys-curve-fCS}) gives $r=2e^{-1}$. 
After $\theta$ reaches $\theta_*$, $\theta$ starts decreasing until $\pi$ while $r$ keeps decreasing until $0$. 
Then, the physical path approaches $\gamma=0$, as in the right of Figure \ref{physical path}. 
We have found in subsection \ref{local-inf-fCS} that $t$ diverges at $\gamma=0$. 
Therefore, we find that the regular singularity at $\gamma=0$ corresponds to the strong coupling limit for the case $\nu_0>4e^{-1}$. 
Note that $W_1$ behaves as ${\cal O}(t^{-1})$, with $tW_1$ approaches $2(\nu+2)$, in contrast to the case $\nu_0<4e^{-1}$ in which $tW_1$ varies periodically. 
\end{itemize}

Let us take a closer look at the transition point $\nu_0=4e^{-1}$. 
For this case, we find $\theta_*=\frac32\pi$ which corresponds to 
\begin{equation}
\gamma_*\ =\ \frac2ee^{3\pi i/2}\ =\ -\frac\nu2. 
\end{equation}
This indicates that the regular singularity at $\gamma=-\frac\nu2$ lies on the physical path for $\nu_0=4e^{-1}$. 
Around $\gamma=\gamma_*$, $t$ behaves as 
\begin{equation}
t(\gamma_*+\delta)\ =\ t(\gamma_*)+\frac{\nu+2\gamma_*}{\gamma_*^2}\delta-\frac{\nu+\gamma_*}{\gamma_*^3}\delta^2+{\cal O}(\delta^3), \hspace{1cm} t(\gamma_*)\ =\ \left( 3\pi+\frac2e \right)i. 
\end{equation}
Since the second term in the right-hand side vanishes, the physical path bifurcates at $\gamma=\gamma_*$. 
This can be understood by noticing that, for the case $\nu_0<4e^{-1}$, there is another path starting from and arriving at $\gamma=0$ which is disconnected from the physical path studied above. 
The singularity at $\gamma=-\frac\nu2$ lies between these two paths. 
At the transition point $\nu_0=4e^{-1}$, these two paths intersect at the singularity, and then they recombine to form a single path for $\nu_0>4e^{-1}$. 
Note that this transition occurs after the supersymmetry is spontaneously broken, and its physical relevance is unclear. 

There are two more singularities. 
At $\gamma=\infty$, we find that ${\rm Re}(t)$ diverges, and therefore, this singularity is always away from the physical path. 
The last singularity at $\gamma=-\frac8\nu$ may collide with the physical path, but it is not clear whether some interesting phenomenon occurs. 

\vspace{5mm}

Next, we consider ABJ theory. 
It is argued in \cite{Aharony:2008gk} that ABJ theory is a consistent unitary theory as long as 
\begin{equation}
N_2-N_1\ \le \ k
\end{equation}
is satisfied, where we assume $N_2\ge N_1$ and $k>0$. 
When the equality holds, we obtain 
\begin{equation}
t_1-t_2\ =\ -2\pi i, 
   \label{boundary-ABJ}
\end{equation}
which corresponds to $B=-2\pi i$. 
This implies that the physically relevant values of $B$ satisfy ${\rm Re}(B)=0$ and $-2\pi\le{\rm Im}(B)\le2\pi$. 

The equality (\ref{boundary-ABJ}) can be realized at the singularity $\gamma=2-2e^B$. 
At this singularity, $t_1$ and $W^{(1)}_1$ vanish, while we find 
\begin{equation}
t_2\ =\ -B, \hspace{1cm} W^{(2)}_1\ =\ \frac{c_2}4\frac{1-e^B}B. 
\end{equation}
This shows that $W^{(2)}_1$ vanishes at $B=-2\pi i$, similar to CS theory at the supersymmetry breaking point. 
Indeed, these two situations are deeply related. 
Since $t_1=0$ implies that there are no matters in the bi-fundamental representation, ABJ theory just reduces to CS theory. 
The comparison with the Wilson loop in CS theory confirms $c_2=-4$. 

The analysis of the physical path for ABJ theory is more difficult than that for fCS theory since we do not have a concise formula for the 't~Hooft couplings. 
Instead, we analyze the physical path around the singularities $\gamma=2-2e^B$ and $\gamma=\infty$ by using the local solutions obtained in section \ref{local solutions}. 
Based on these solutions, we try to guess the global property of the physical path. 
Note that the physical path for ABJM theory connects these two singularities \cite{Marino:2009jd}. 

The local solution for $t_1$ around $\gamma=2-2e^B$ is 
\begin{equation}
t_1\ =\ \frac14e^{-B/2}\xi+{\cal O}(\xi^2), \hspace{1cm} \xi\ =\ \gamma-(2-2e^B). 
\end{equation}
For $B=0$, the physical path corresponds to $\xi\in i\mathbb R$ when $\xi$ is small. 
Suppose that we shift $B$ by $i\alpha$ with a real $\alpha$. 
Then, in order to keep the condition ${\rm Re}(t_1)=0$, we have to choose the values of $\xi$ from $ie^{i\alpha/2}\mathbb R$. 
That is, the physical path near $\xi=0$ must be rotated by the angle $\alpha/2$. 

On the other hand, $t_1$ around $\gamma=\infty$ behaves as 
\begin{equation}
t_1\ =\ \frac{c_3}2(\log\gamma)^2-\frac12\Bigl[ c_2+c_3(B-2)+2\tilde c_3 \Bigr]\log\gamma+{\cal O}(\gamma^0) 
\end{equation}
with $c_3=2i/\pi$. 
Up to this order, we find that the shift of $B$ by $i\alpha$ can be canceled again by rotating $\gamma$ by $e^{i\alpha/2}$. 
Therefore, the physical path near infinity is also rotated by exactly the same angle as the physical path around $\gamma=2-2e^B$. 
From this analysis, it is natural to expect that the physical path of ABJ theory would be rotated globally by the shift of $B$. 
We do not see any sign of a phase transition in this case. 

\vspace{1cm}

\section{Recursion relation}   \label{recursion relation}

\vspace{5mm}

So far, we have discussed the quantities $\pi_0,\pi_1,\pi_2$. 
It is possible to derive a set of similar differential equations for $\pi_n$ with $n\ge3$. 
However, the solutions of them would contain new integration constants which must be fixed independently. 
Instead, as mentioned in \cite{Suyama:2025ydf}, $\pi_n$ can be determined from $\pi_0,\pi_1,\pi_2$ via a recursion relation. 
In this section, we will make use of the recursion relation to determine the $n$-dependence of $\pi_n$. 
The $t$-dependence of $\pi_n$ is then determined by using the common $t$ obtained in the previous section. 
Appendix \ref{app-WL} gives the expressions for the multiply-wound Wilson loops $W_n$ in terms of $\pi_n$. 

\vspace{5mm}

The recursion relation \cite{Suyama:2025ydf} can be derived as follows. 
We multiply $x^{n+1}f(x)$ to the both sides of (\ref{x-derivative}), where $f(x)$ was defined in (\ref{f_i,f}). 
Then we obtain 
\begin{equation}
2\partial_x\Bigl( x^{n+1}f(x) \Bigr)y+x^{n+1}p(x)y\ =\ \partial_x(\cdots), 
   \label{before-int}
\end{equation}
where 
\begin{equation}
p(x)\ =\ f(x)\sum_{i=1}^4\frac{l_i}{x-\alpha_i}
\end{equation}
is a polynomial of degree 2 since we assume $\sum_{i=1}^4l_i=0$ as before. 
By integrating both sides of (\ref{before-int}), we obtain a recursion relation among $\pi_n$. 

In the following, we concentrate on ABJ theory. 
Then we obtain 
\begin{eqnarray}
& & (2n+10)\pi_{n+4}-(2n+9)(\alpha_3+\alpha_4)\pi_{n+3}+\Bigl[ (2n+8)\alpha_3\alpha_4-4(2n+4) \Bigr]\pi_{n+2} \nonumber \\ [1mm] 
& & +4(2n+3)(\alpha_3+\alpha_4)\pi_{n+1}-4(2n+2)\alpha_3\alpha_4\pi_n\ =\ 0
   \label{recursion}
\end{eqnarray}
for $n\ge0$. 
Note that this is also valid for $n=-1$ for which we find 
\begin{equation}
8\pi_3-7(\alpha_3+\alpha_4)\pi_2+(6\alpha_3\alpha_4-8)\pi_1+4(\alpha_3+\alpha_4)\pi_0\ =\ 0. 
   \label{recursion-pi3}
\end{equation}
Therefore, once $\pi_0,\pi_1,\pi_2$ are determined by the PF equation (\ref{PF-WL}), we can also determine $\pi_n$ recursively for any $n\ge3$. 

It is rather difficult, however, to analyze the five-term relation (\ref{recursion}) obtained above. 
To simplify the analysis, we introduce the generating function 
\begin{equation}
g(t)\ :=\ \sum_{n=0}^\infty \pi_nt^{n+1}. 
   \label{generating fn}
\end{equation}
Then, the recursion relation (\ref{recursion}) can be rewritten as a differential equation for $g(t)$ as follows. 

The relation (\ref{before-int}) multiplied by $t^{n+4}$ can be written as 
\begin{equation}
2\sum_{m=0}^4d_mt^{4-m}\partial_t\Bigl[ x^{n+m}yt^{n+m+1} \Bigr]+\sum_{m=0}^2\tilde d_mt^{2-m}x^{n+m+1}yt^{n+m+2}\ =\ \partial_x(\cdots), 
\end{equation}
where $d_m$ and $\tilde d_m$ are defined by 
\begin{equation}
f(x)\ =\ \sum_{m=0}^4d_mx^m, \hspace{1cm} p(x)\ =\ \sum_{m=0}^2\tilde d_mx^m. 
\end{equation}
Then, the summation on $n$ gives us a first-order linear differential equation 
\begin{eqnarray}
2t^4f(t^{-1})\partial_tg(t)+t^2p(t^{-1})g(t) 
&=& \Bigl[ 6\pi_2-5(\alpha_3+\alpha_4)\pi_1+4\alpha_3\alpha_4\pi_0 \Bigr]t^2 \nonumber \\ [1mm] 
& & +\Bigl[ 4\pi_1-3(\alpha_3+\alpha_4)\pi_0 \Bigr]t+2\pi_0 
   \label{DE-g(t)}
\end{eqnarray}
for $g(t)$, where we have used the relation (\ref{recursion-pi3}) to eliminate $\pi_3$ in the right-hand side. 
The integration constant for $g(t)$ is fixed by $g(0)=0$ required by the definition (\ref{generating fn}). 
The right-hand side can be written more explicitly as 
\begin{equation}
\Bigl[ 6\pi_2-10\gamma\pi_1+(4\gamma^2-16e^{2B})\pi_0 \Bigr]t^2+\Bigl[ 4\pi_1-6\gamma\pi_0 \Bigr]t+2\pi_0. 
   \label{RHS-ABJ}
\end{equation}
The exact solution of (\ref{DE-g(t)}) seems to be quite complicated. 
Instead, we will solve this equation perturbatively in terms of the local coordinate $\xi$ around some of the  regular singularities. 

First, we consider the singularity at $\gamma=2-2e^B$ whose local coordinate is $\xi=\gamma-(2-2e^B)$. 
For the local solution, we choose (\ref{local-ABJ-WL}) which vanishes at the singularity. 
Then, the right-hand side (\ref{RHS-ABJ}) becomes 
\begin{equation}
-\left[ 8e^{B/2}t^2+4e^{-B/2}(3e^B-1)t+2e^{-B/2} \right]\xi+{\cal O}(\xi^2).  
\end{equation}
This suggests that we should expand the generating function as 
\begin{equation}
g(t)\ =\ \sum_{m=0}^\infty g_m(t)\xi^{m+1}. 
\end{equation}
We find that $g_0(t)$ satisfies 
\begin{eqnarray}
& & 2(2t-1)^2(2t+1)\Bigl[ (4e^B-2)t+1 \Bigr]\partial_tg_0(t)+4(e^B-1)(2t-1)^2g_0(t) \nonumber \\ [1mm] 
&=& -8e^{B/2}t^2-4e^{-B/2}(3e^B-1)t-2e^{-B/2}. 
\end{eqnarray}
The general solution 
\begin{equation}
g_0(t)\ =\ c\sqrt{\frac{1+2t}{1+(4e^B-2)t}}-e^{-B/2}\frac t{1-2t} 
\end{equation}
is remarkably simple. 
The requirement $g(0)=0$ imposes $c=0$. 
The expansion of $g_0(t)$ in terms of $t$ gives 
\begin{equation}
g_0(t)\ =\ -e^{-B/2}\sum_{n=0}^\infty 2^nt^{n+1}. 
\end{equation}
Therefore, we find that $\pi_n$ behave as 
\begin{equation}
\pi_n\ =\ -2^ne^{-B/2}\xi+{\cal O}(\xi^2). 
\end{equation}
Note that these are exact for $n$. 
We can check that these leading terms satisfy the recursion relation (\ref{recursion}).  
For large $n$ and small $\xi$, the recursion relation (\ref{recursion}) becomes 
\begin{equation}
2n\pi_{n+4}+8(e^B-1)n\pi_{n+3}-16e^Bn\pi_{n+2}+32(1-e^B)n\pi_{n+1}+32(2e^B-1)n\pi_n\ \sim\ 0. 
\end{equation}
This is indeed satisfied by the leading terms of $\pi_n$. 

Next, we consider the singularity at $\gamma=\infty$. 
In terms of the local coordinate $\xi=\gamma^{-1}$, we find 
\begin{equation}
f(t^{-1})\ =\ {\cal O}(\xi^{-2}), \hspace{1cm} p(t^{-1})\ =\ {\cal O}(\xi^{-2})
%, \hspace{1cm} {\rm RHS}\ =\ {\cal O}(\xi^{-3}), 
\end{equation}
while (\ref{RHS-ABJ}) is of order ${\cal O}(\xi^{-3})$, where we used the estimates $\pi_n={\cal O}(\xi^{-n-1})$ for $n\le2$ obtained from the analysis of the PF equation (\ref{PF-WL}). 
This looks to suggest that we should assume $g(t)={\cal O}(\xi^{-1})$, but this is not compatible with the definition of $g(t)$. 
Therefore, a straightforward expansion does not work. 

In fact, the appearance of this wrong result could be anticipated by examining the coefficient function of $\partial_tg(t)$ which is 
\begin{equation}
t^4f(t^{-1})\ =\ (1-2t)(1+2t)(1-\alpha_3t)(1-\alpha_4t). 
\end{equation}
Since both $\alpha_3$ and $\alpha_4$ diverge in the limit $\gamma\to\infty$, two of the singularities of the differential equation (\ref{DE-g(t)}), namely $t=\alpha_3^{-1},\alpha_4^{-1}$, approach the origin. 
Then, the determination of a power series solution for $g(t)$ around $t=0$ would be subtle. 
Another explanation of this phenomenon is as follows. 
Since $g(t)$ is defined as (\ref{generating fn}), we can obtain $\pi_n$ by a contour integration of $t^{-n-2}g(t)$ whose contour encircles $t=0$ but excludes all the singularities of $g(t)$. 
When a singularity, say $t=\alpha_3^{-1}$, approaches the origin, then the contour must be deformed such that it avoids this singularity. 
Therefore, the behavior of $g(t)$ for small $t$ is also relevant for determining $\pi_n$. 
In this case, we cannot assume $t={\cal O}(\xi^0)$ in the expansion of (\ref{DE-g(t)}). 

To remedy this situation, we rescale the variable $t$. 
In terms of $s:=\xi^{-1} t$, no singularity approaches the origin in the limit $\xi\to0$, and we safely expand the equation (\ref{DE-g(t)}) with the assumption $s={\cal O}(\xi^0)$. 
We also expand the generating function as 
\begin{equation}
g(t)\ =\ \sum_{n=0}^\infty g_n(s)\xi^n, 
\end{equation}
where we allow $g_n(s)$ to contain $\log \xi$. 
Then, the leading contribution $g_0(s)$ satisfies 
\begin{equation}
(s-1)^2\partial_sg_0(s)+(s-1)g_0(s)\ =\ C(s-1)+c_3, 
\end{equation}
where 
\begin{equation}
C \ =\ -c_2-\frac{c_3(2B-3)}2-c_3\log\xi. 
\end{equation}
The solution satisfying $g_0(0)=0$ turns out to be 
\begin{eqnarray}
g_0(s) 
&=& -C\frac s{1-s}-c_3\frac{\log(1-s)}{1-s} \nonumber \\ [1mm] 
&=& \sum_{n=0}^\infty \left[ -C+c_3\sum_{m=1}^{n+1}\frac1m \right]s^{n+1}. 
\end{eqnarray}
Therefore, we find that $\pi_n$ for large $n$ behave as 
\begin{equation}
\pi_n\ \sim\ \left[ -C+c_3\sum_{m=1}^{n+1}\frac1m \right]\xi^{-n-1}\ \sim\ c_3\log (n+1)\,\xi^{-n-1}. 
\end{equation}
Indeed, these satisfy the recursion relation (\ref{recursion}) for large $n$ and small $\xi$ since 
\begin{equation}
(2n+10)\log(n+5)-2(2n+9)\log(n+4)+(2n+8)\log(n+3)\ =\ -\frac1{n^2}+{\cal O}(n^{-3}) 
\end{equation}
holds. 

\vspace{1cm}

\section{Discussion}   \label{discuss}

\vspace{5mm}

We have investigated the PF equations for Wilson loops and 't~Hooft coupling for fCS and ABJ theories. 
We found concise expressions (\ref{A})(\ref{tilde A}) for the coefficient matrices of the equations. 
These expressions enable us to understand the singularity structure of the equations. 
For example, we found that most of the singularities are simple poles. 
In addition, there are double poles, for which we found explicit gauge transformations which eliminate them. 
Then, we solved the equations locally around the singularities, and extracted information on physics of fCS and ABJ theories. 
In contrast to the previous work \cite{Suyama:2025ydf}, the local solutions depend on a parameter of the theory. 
The dependence on the parameter can be determined by another PF equations with respect to the parameter. 
For fCS theory, we found an interesting phase transition as we vary the parameter $\nu$. 
We also analyzed the multiply-wound Wilson loops $W_n$ by employing the recursion relation obtained in \cite{Suyama:2025ydf}, and derived the $n$-dependence of them. 

\vspace{5mm}

An advantage of the analysis based on PF equations is that the physically relevant information of Wilson loops is concentrated on the singularities of the equations. 
We have observed in section \ref{local solutions} that the leading behavior of the Wilson loop $W_1$ is given by four exponents $\delta_1,\cdots,\delta_4$ defined in (\ref{exponents 1})(\ref{exponents 2}). 
For example, an exponential growth of $W_1$ is possible only when $\delta_1=0$ is satisfied. 
These exponents are largely determined by the residue matrices of the corresponding singularity. 
Therefore, the investigation on the possible behaviors of $W_1$ in a given theory may be reduced to the classification of the possible singularity structure of the PF equations. 

We have also observed in section \ref{singularities} that the singularity structure is closely related to various degenerations of the algebraic curve (\ref{general curve}). 
For example, at least for the cases we investigated, a more complicated degeneration of the algebraic curve leads to a higher order pole in the PF equations. 
It is quite interesting if the singularity structure of the PF equations including the residue matrices can be directly obtained from the singularity structure appearing on the algebraic curve. 
This would give us a geometric understanding on the various limiting behaviors of $W_1$ in a given Chern-Simons-matter theory. 
It is quite possible that related results are already known in mathematics literature. 

We have succeeded in determining the parameter dependence of the local solutions, but they still depend on a few unknown constants. 
For fCS and ABJ theories, these constants can be also determined in many cases since we know the solutions at a particular value of the parameter. 
Even without such a knowledge, it would be possible to find at least the qualitative behavior of the Wilson loops. 
As observed in section \ref{local solutions}, the exact results for fCS theory exhibit that the leading behavior of the physical solution is given by the leading term of the local solution with generic values of those constants. 
It is interesting to check whether this is a generic situation. 
Another possibility is to investigate the integral expressions (\ref{integral-WL})(\ref{integral-tH}) in more detail. 
Since $y(x)$ is known from (\ref{general curve}), these expressions contain all the information on the Wilson loops. 
Therefore, if we would understand the correct choice of the integration contour $C$ for a chosen singularity, then we could determine the constants in principle by suitably expanding the integrals. 
This is actually complementary to the analysis of PF equations. 
The integral expressions contain global information, while the PF equations are more useful for extracting local information. 

\vspace{5mm}

The application of PF equations can be further extended to more general Chern-Simons-matter theories. 
For example, we can consider GT theory \cite{Gaiotto:2009mv}\cite{Gaiotto:2009yz} which has the gauge group ${\rm U}(N_1)\times{\rm U}(N_2)$ with a generic Chern-Simons levels $k_1$ and $k_2$. 
This reduces to ABJ theory when $k_1+k_2=0$ is satisfied. 
In the context to AdS/CFT correspondence \cite{Gaiotto:2009mv}\cite{Fujita:2009kw}, the combination $k_1+k_2$ corresponds to the Romans mass \cite{Romans:1985tz}, or in other words, an R-R flux \cite{Polchinski:1995mt}. 
Therefore, GT theory is holographically dual to massive Type IIA theory. 
It is known that GT theory has qualitatively different behaviors of the free energy \cite{Jafferis:2011zi} and the Wilson loop \cite{Suyama:2011yz} from those of ABJ theory. 
It is interesting to understand how those information can be encoded into the corresponding PF equations. 

We also expect that a similar research can be performed for theories with many parameters. 
In general, we will have one differential equation for each parameter which can be used to determine the parameter dependence completely, up to a few constants. 
In practice, it will be important to choose a good parametrization of the corresponding algebraic curve so that the analysis becomes simpler. 
For example, if a given theory has gauge group $\prod_i{\rm U}(N_i)$, then we have one 't~Hooft coupling for each $i$. 
In this case, it is not clear a priori which combination of the 't~Hooft couplings we should choose as the variable $\gamma$ for PF equations. 
If the theory has a gravity dual, then a natural choice would be the one corresponding to the radius of AdS space. 
It is interesting if there is a criterion for choosing good coordinates in the parameter space of the theory purely from the field theory point of view. 

\appendix

\vspace{1cm}

\section{Algebraic curve of fCS theory}   \label{app-fCS}

\vspace{5mm}

We derive the algebraic curve (\ref{fCS curve}) from the localization formula 
\begin{equation}
Z\ =\ \int du\,\exp\left( \sum_{i=1}^N\frac{ik}{4\pi}u_i^2 \right)\frac{\displaystyle \prod_{i<j}\sinh^2\frac{u_i-u_j}2}{\displaystyle \prod_{i=1}^N\left( \cosh\frac{u_i}2 \right)^{N_f}}
\end{equation}
for the partition function of fCS theory. 
The saddle points of the integral satisfy 
\begin{equation}
\frac k{2\pi i}u_i-\sum_{j\ne i}\coth\frac{u_i-u_j}2+\frac{N_f}2\tanh\frac{u_i}2\ =\ 0. 
   \label{saddle-point}
\end{equation}

\begin{figure}
\begin{center}
\includegraphics[width=70mm]{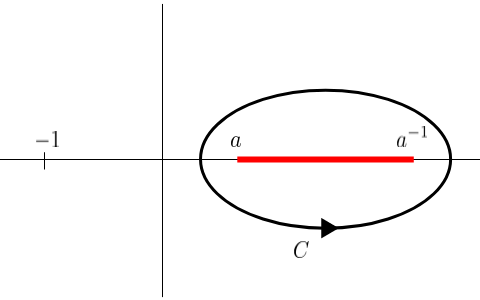}
\end{center}
\caption{
The branch cut of the planar resolvent $v(z)$. 
The integration contour $C$ for (\ref{z-integral-WL})(\ref{integral-tH}) is also depicted. 
Note that $a$ for the branch point may take complex values. 
}
   \label{branch cut}
\end{figure}

We define the resolvent 
\begin{equation}
v(z)\ :=\ \frac tN\sum_{i=1}^N\frac{z+z_i}{z-z_i}, \hspace{1cm} z_i\ :=\ e^{u_i}, 
\end{equation}
where $t$ is the 't~Hooft coupling (\ref{parameters-fCS}). 
In the planar limit, $v(z)$ becomes a function which has a square-root branch cut on an interval $[a,a^{-1}]$ and is finite at infinity. 
The branch cut is depicted in Figure \ref{branch cut}. 
Note that the distribution of the eigenvalues $u_i$ collapses to a point at the origin in the weak coupling limit $k\to\infty$. 
This corresponds to the limit $a\to1$. 

The saddle-point equations (\ref{saddle-point}) can be rewritten in terms of $v(z)$ as 
\begin{equation}
2+\frac{2\nu x}{(x+1)^2}\ =\ xv'(x_+)+xv'(x_-), 
\end{equation}
where $x$ is a point on the branch cut and $x_\pm$ are two points infinitesimally above and below $x$. 
This can be further simplified by introducing 
\begin{equation}
\omega(z)\ :=\ -1-\frac{\nu z}{(z+1)^2}+zv'(z), 
   \label{omega(z)}
\end{equation}
by which we obtain 
\begin{equation}
\omega(x_+)+\omega(x_-)\ =\ 0. 
\end{equation}
Note that $\omega(z)$ is invariant under the inversion $z\to z^{-1}$, which is due to the invariance of (\ref{saddle-point}) under $u_i\to-u_i$. 

We find that $\omega(z)$ has the following analytic structure: 
\begin{itemize}
\item 
$\omega(z)^2$ has simple poles at $z=a,a^{-1}$, 
\item 
$\omega(z)^2$ has a quadruple pole at $z=-1$, 
\item 
$\omega(\infty)=-1$, 
\item 
$\omega(z)$ has only one branch cut. 
\end{itemize}
These conditions with the inversion symmetry almost determine $\omega(z)^2$ as 
\begin{equation}
\omega(z)^2\ =\ \frac{(z-1)^2(z^2-(A-1)z+1)^2}{(z-a)(z-a^{-1})(z+1)^4}. 
   \label{omega^2}
\end{equation}
The remaining constant $A$ can be fixed by requiring that this gives the correct coefficient of the double pole given by (\ref{omega(z)}). 
We find 
\begin{equation}
A\ =\ -1-\frac\nu2\left( a^{1/2}+a^{-1/2} \right). 
\end{equation}
In terms of a new variable $x:=z+z^{-1}$, the right-hand side of (\ref{omega^2}) can be rewritten as 
\begin{equation}
\frac{(x-2)(x-(A-1))^2}{(x-(a+a^{-1}))(x+2)^2}. 
\end{equation}

We introduce 
\begin{equation}
\gamma\ :=\ a^{1/2}+a^{-1/2}. 
\end{equation}
In terms of $\gamma$, we find 
\begin{equation}
\alpha\ :=\ a+a^{-1}\ =\ \gamma^2-2, \hspace{1cm} \beta\ :=\ A-1\ =\ -\frac\nu2\gamma-2. 
\end{equation}
Now we find that $\omega(z)$ is a solution of 
\begin{equation}
y^2\ =\ \frac{(x-2)(x-\beta)^2}{(x+2)^2(x-\alpha)}, 
\end{equation}
which defines the algebraic curve of fCS theory.

\vspace{1cm}

\section{A formula for Wilson loops}   \label{app-WL}

\vspace{5mm}

We derive a formula which expresses the Wilson loops $tW_n$ given in (\ref{z-integral-WL}) as a linear combination of $\pi_n$ given in (\ref{integral-WL}). 

We start with 
\begin{equation}
tW_{n+1}\ =\ -\frac1{2(n+1)}\int_C\frac{dz}{2\pi i}z^ny(x), \hspace{1cm} x\ =\ z+z^{-1}. 
\end{equation}
This can be rewritten as 
\begin{eqnarray}
-\frac1{2(n+1)}\int_C\frac{dz}{2\pi i}z^ny(x) 
%&=& -\frac1{4(n+1)\pi i}\int_C\frac{dz}zz^{z+1}y(x) \nonumber \\ 
%&=& -\frac1{8(n+1)\pi i}\int_C\frac{dz}z\left( z^{z+1}-\frac1{z^{n+1}} \right)y(x) \nonumber \\
&=& -\frac1{8(n+1)\pi i}\int_C\frac{dz}z\left( z-\frac1z \right)\frac{z^{n+1}-z^{-n-1}}{z-z^{-1}}y(x) \nonumber \\ [1mm] 
%&=& -\frac1{4(n+1)}\int_C\frac{dz}{2\pi i}\left( 1-\frac1{z^2} \right)U_n\left( \frac{z+z^{-1}}2 \right)y(x) \nonumber \\
&=& -\frac1{4(n+1)}\int_C\frac{dx}{2\pi i}U_n(x/2)y(x), 
\end{eqnarray}
where $U_n(x)$ are the Chebyshev polynomials of the second kind. 
For example, 
\begin{equation}
U_0(x)\ =\ 1, \hspace{1cm} U_1(x)\ =\ 2x, \hspace{1cm} U_2(x)\ =\ 4x^2-1. 
\end{equation}
In general, they are written as 
\begin{equation}
U_n(x)\ =\ \sum_{k=0}^nu_kx^k. 
\end{equation}
Using the coefficients $u_k$, $tW_{n+1}$ can be written as 
\begin{equation}
tW_{n+1}\ =\ -\frac1{4(n+1)}\sum_{k=0}^n\frac{u_k}{2^k}\pi_k. 
   \label{pi to W}
\end{equation}
This implies, for example, 
\begin{equation}
W_1\ =\ -\frac1{4t}\pi_0, \hspace{1cm} W_2\ =\ -\frac1{8t}\pi_1, \hspace{1cm} W_3\ =\ -\frac1{12t}\left( \pi_2-\pi_0 \right). 
\end{equation}

\vspace{1cm}

\section{Exact results for fCS theory}   \label{app-exact}

\vspace{5mm}

In the planar limit, the $n$-times-wound Wilson loops are given as \cite{Kapustin:2009kz} 
\begin{equation}
W_n\ =\ \frac1N\sum_{i=1}^Ne^{nu_i}, 
\end{equation}
where $u_i$ satisfy the saddle-point equations (\ref{saddle-point}). 
These are encoded into the resolvent $v(z)$ as 
\begin{equation}
v(z)\ =\ t+2\sum_{n=1}^\infty tW_nz^{-n}. 
\end{equation}
In appendix \ref{app-fCS}, we defined $\omega(z)$ from $v(z)$ as in (\ref{omega(z)}). 
This can be expanded as 
\begin{equation}
\omega(z)\ =\ -\frac{(z-1)(z^2-\beta z+1)}{(z+1)^2\sqrt{(z-a)(z-a^{-1})}}\ =\ -1+\sum_{n=1}^\infty \omega_nz^{-n}. 
\end{equation}
This implies the exact formulas 
\begin{equation}
-2ntW_n\ =\ \omega_n-(-1)^nn\nu. 
\end{equation}

In section \ref{local solutions}, we need the exact results for $\pi_n$. 
The relation between $W_n$ and $\pi_n$ is given as (\ref{pi to W}) in appendix \ref{app-WL}. 
By inverting this relation, we obtain, for example, 
\begin{equation}
\left[
\begin{array}{c}
\pi_0 \\
\pi_1 \\
\pi_2
\end{array}
\right] 
\ =\ 
\left[
\begin{array}{l}
-\gamma^2-\nu\gamma+2\nu+4 \\ [1mm] 
-\frac34\gamma^4-\frac\nu2\gamma^3+4\gamma^2+4\nu\gamma-4\nu-4 \\ [1mm] 
-\frac58\gamma^6-\frac{3\nu}8\gamma^5+\frac92\gamma^4+3\nu\gamma^3-10\gamma^2-10\nu\gamma+8\nu+8
\end{array}
\right]. 
   \label{exact WL-fCS}
\end{equation}
Indeed, this satisfies the PF equation for Wilson loops (\ref{PF-WL}). 

\begin{figure}
\begin{center}
\includegraphics[width=70mm]{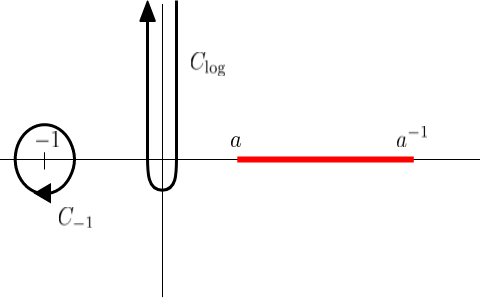}
\end{center}
\caption{
The contours $C_{\log}$ and $C_{-1}$ obtained by deforming the contour $C$ in Figure \ref{branch cut}. 
}
   \label{deformed contour}
\end{figure}

The exact formula for the 't~Hooft coupling $t$ can be obtained by performing the integral of (\ref{integral-tH}). 
For this purpose, we first rewrite the integral as 
\begin{equation}
t\ =\ -\frac1{4\pi i}\int_C\frac{dz}z\log z\left[ 1-\frac{(z-1)(z^2-\beta z+1)}{(z+1)^2\sqrt{(z-a)(z-a^{-1})}} \right]
\end{equation}
which improves the behavior of the integrand at infinity. 
The integration contour $C$ encircles the branch cut counterclockwise, as depicted in Figure \ref{branch cut}. 
We deform the contour as depicted in Figure \ref{deformed contour}, where we choose the branch cut of $\log z$ to lie on the positive imaginary axis. 
The integral along $C_{-1}$ gives $\frac\nu2$. 
The other integral along $C_{\rm log}$ can be written as 
\begin{equation}
\frac12\int_0^{i\infty}\frac{dz}z\left[ 1-\frac{(z-1)(z^2-\beta z+1)}{(z+1)^2\sqrt{(z-a)(z-a^{-1})}} \right]\ =\ 2\log\frac\gamma2-\frac\nu\gamma. 
\end{equation}
Then, we obtain 
\begin{equation}
t\ =\ 2\log\frac\gamma2+\frac\nu2\left( 1-\frac2\gamma \right). 
   \label{exact tH-fCS}
\end{equation}
$\tau_n$ for $n\ge1$ can be also obtained similarly.

\end{document}